\documentclass[traditabstract,epsf]{aa}
\usepackage{epsfig}
\usepackage{txfonts}

\def\ms{\,m\,s$^{-1}$\,}         
\def\kms{\,km\,s$^{-1}$\,}         
\def\m2s2{\hbox{\,m$^{2}$\,s$^{-2}$}} 

\begin{document}
\title{The TRAPPIST survey of southern transiting planets. I}
\subtitle{Thirty eclipses of the ultra-short period planet WASP-43\,b\thanks{Based on data collected with the TRAPPIST and {\it Euler} telescopes at ESO 
La Silla Observatory, Chile, and with the VLT/HAWK-I instrument at ESO Paranal Observatory, 
Chile (program 086.C-0222).}}

\author{ 
M.~Gillon\inst{1}, A.~H.~M.~J.~Triaud\inst{2}, J.~J.~Fortney\inst{3}, B.-O.~Demory\inst{4}, E.~Jehin\inst{1}, 
M.~Lendl\inst{2}, P.~Magain\inst{1}, P.~Kabath\inst{5}, D.~Queloz\inst{2}, R.~Alonso\inst{2}, D.~R.~Anderson\inst{6}, 
A.~Collier~Cameron\inst{7}, A.~Fumel\inst{1}, L.~Hebb\inst{8}, C.~Hellier\inst{6},  A.~Lanotte\inst{1}, 
P.~F.~L.~Maxted\inst{6}, N.~Mowlavi\inst{2}, B.~Smalley\inst{6}}

\offprints{michael.gillon@ulg.ac.be}
\institute{
        $^1$ Universit\'e de Li\`ege, All\'ee du 6 ao\^ut 17, Sart Tilman, Li\`ege 1, Belgium\\
        $^2$ Observatoire de Gen\`eve, Universit\'e de Gen\`eve, 51 Chemin des Maillettes, 1290 Sauverny, Switzerland\\
        $^3$ Department of Astronomy and Astrophysics, University of California, Santa Cruz, CA 95064, USA\\
        	$^4$ Department of Earth, Atmospheric and Planetary Sciences, Department of Physics, Massachusetts Institute of Technology, 77 Massachusetts Ave., Cambridge, MA 02139, USA\\
	$^5$ European Southern Observatory, Alonso de Cordova 3107, Casilla 19001, Santiago, Chile\\
	$^6$ Astrophysics Group, Keele University, Staffordshire ST5 5BG, UK\\
	$^7$ School of Physics and Astronomy, University of St. Andrews, North Haugh, Fife, KY16 9SS, UK\\
	$^8$ Department of Physics and Astronomy, Vanderbilt University, Nashville, TN 37235, USA\\
	}

\date{Received date / accepted date}
\authorrunning{M. Gillon et al.}
\titlerunning{Thirty eclipses of the ultra-short period planet WASP-43\,b}
\abstract{We present twenty-three transit light curves and seven occultation light curves for the ultra-short 
period planet WASP-43\,b, in addition to eight new measurements of the radial velocity of the star.
Thanks to this extensive data set, we improve significantly the parameters of the system. Notably, 
the largely improved precision on the stellar density ($2.41 \pm 0.08$ $\rho_\odot$) combined with 
constraining the age to be younger than a Hubble time allows us to break the degeneracy of
 the stellar solution mentioned in the discovery paper. The resulting stellar mass and size are 
 $0.717 \pm 0.025$ $M_\odot$ and $0.667 \pm 0.011$ $R_\odot$. Our deduced physical parameters for 
 the planet are $2.034 \pm 0.052$ $M_{Jup}$ and $1.036 \pm 0.019$ $R_{Jup}$. Taking into account its 
 level of irradiation, the high density of the planet favors an old age and a massive core. Our deduced 
 orbital eccentricity, $0.0035_{-0.0025}^{+0.0060}$, is consistent with a fully circularized orbit.
We detect the emission of the planet at 2.09 $\mu$m at better than 11-$\sigma$, the deduced occultation
depth being $1560 \pm 140$ ppm. Our detection of the occultation at 1.19 $\mu$m is marginal ($790 \pm 320$ ppm) and more observations are needed to confirm it. We place a 3-$\sigma$ upper limit of 850 ppm on the depth of the occultation at $\sim$0.9 $\mu$m. Together, these results strongly favor a poor  redistribution of the heat to the night-side of the planet, and marginally favor a model with no day-side temperature inversion. 

}

\maketitle

\section{Introduction}

There are now more than seven hundreds planets known to orbit around other stars than our Sun
(Schneider 2011). A significant fraction of them are Jovian-type planets orbiting within 0.1 AU of their 
host stars. Their very existence poses an interesting challenge for our theories of planetary formation 
and evolution, as such massive planets could not have formed so close to their star (D'Angelo et al. 
2010; Lubow \& Ida 2010). Long neglected in favor of disk-driven migration theories (Lin et al. 1996), 
the postulate that past dynamical interactions combined with tidal dissipation could have shaped 
their present orbit has raised a lot of interest recently (e.g. Fabrycky \& Tremaine 2007, Naoz et al. 
2011, Wu \& Lithwick 2011), partially thanks to the discovery that a significant fraction of these planets 
have high orbital obliquities that suggest past violent dynamical perturbations (e.g. Triaud et al. 2010).
 
Not only the origin of these planets but also their fate raises many questions. Their relatively large 
mass and small semi-major axis imply tidal interactions with the host star that should lead in most cases 
to a slow spiral-in of the planet and to a transfer of angular momentum to the star (e.g. Barker \& Ogilvie 
2009, Jackson et al. 2009, Mastumura et al. 2010), the end result being the planet's disruption  at its
Roche limit (Gu et al. 2003). The timescale of this final disruption depends mostly on the 
timescale of the migration mechanism, the tidal dissipation efficiency of both bodies, and the efficiency 
of angular momentum losses from the system due to magnetic braking. These three parameters being
poorly known, it is desirable to detect and study in depth `extreme' hot Jupiters, i.e. massive planets 
having exceptionally short semi-major axes that could be in the final stages of their tidal orbital decay.

The WASP transit survey detected two extreme examples of such ultra-short period planets, 
WASP-18\,b (Hellier et al. 2009) and WASP-19\,b (Hebb et al. 2010), both having an orbital period 
smaller than one day. Thanks to their transiting nature, these systems were amenable for a thorough 
characterization that made possible a study of the effects of their tidal interactions (Brown et al. 2011). 
Notably, photometric observations of some of their occultations made possible not only to probe both 
planets' dayside emission spectra but also to bring strong constraints on their orbital eccentricity 
(Anderson et al. 2010, Gibson et al. 2010, Nymeyer et al. 2011), a key parameter to assess their tidal 
history and energy budget. The same WASP survey has recently announced the discovery of a third 
ultra-short period Jovian planet called WASP-43\,b (Hellier et al. 2011b, hereafter H11). Its orbital period 
is 0.81 d, the only hot Jupiter having a smaller period being WASP-19\,b (0.79 d). Furthermore, it orbits
around a very cool K7-type dwarf that has the lowest mass among all the stars orbited by a hot Jupiter 
($0.58 \pm 0.05 M_\odot$, $T_{eff}=4400 \pm 200$ K, H11), except for the recently announced M0 dwarf 
KOI-254  ($0.59 \pm 0.06M_\odot$, $T_{eff}=3820 \pm 90$ K, Johnson et al. 2011). Nevertheless, H11
 presented another plausible solution for the stellar mass that is significantly larger, $0.71 \pm 0.05 M_\odot$. 
 This degeneracy translates into a poor knowledge of the physical parameters of the system.  

With the aim to improve the characterization of this interesting ultra-short period planet, we performed an 
intense ground-based photometric monitoring of its eclipses (transits and occultations), complemented with 
new measurements of the radial velocity (RV) of the star. These observations were carried out in the frame 
of a new photometric survey based on the 60cm  robotic telescope TRAPPIST\footnote{see http://www.ati.ulg.ac.be/TRAPPIST} 
({\it TRA}nsiting {\it P}lanets and {\it P}lanetes{\it I}mals {\it S}mall {\it T}elescope; Gillon et al. 2011a, Jehin et al. 2011). 
The concept of this survey is the intense high-precision photometric monitoring of the transits of southern
transiting systems, its goals being (i) to improve the determination of the physical and orbital parameters
of the systems, (ii) to assess the presence of undetected objects in these systems through variability studies of 
the transit parameters, and (iii) to measure or put an upper limit on the  very-near-IR thermal emission of the 
most highly irradiated planets to constrain their atmospheric properties. We complemented the data acquired in the 
frame of this program for the WASP-43 system  by high-precision occultation time-series photometry
gathered in the near-IR with the VLT/HAWK-I instrument (Pirard et al., 2004, Casali et al. 2006) in our program 
086.C-0222. We present here the results of the analysis of this extensive data set. 
Section 2 below presents our  data. Their analysis is described in Sec.~3. We discuss the acquired 
results and drawn inferences about the WASP-43 system in Sec.~4. Finally, we give our conclusions in Sec.~5.
 
\section{Data}

\subsection{TRAPPIST  {\it I+z} filter transit photometry}

We observed 20 transits of WASP-43\,b with  TRAPPIST and its thermoelectrically-cooled 2k $\times$ 2k CCD 
camera with a field of view of 22' $\times$ 22' (pixel scale = 0.65"). All the 20 transits were observed in an Astrodon
`{\it I+z}' filter that has a transmittance $>$90\% from 750 nm to beyond 1100 nm\footnote{see http://www.astrodon.com/products/filters/near-infrared/}, the red end of the effective bandpass being defined by the spectral response of the CCD. This 
wide red filter minimizes the effects of limb-darkening and differential atmospheric extinction while maximizing stellar 
counts. Its effective wavelength for $T_{eff} = 4400 \pm 200$ K is $\lambda_{eff} = 843.5 \pm 1.2$ nm. The 
mean exposure time was 20s. The telescope was slightly defocused to minimize pixel-to-pixel effects and to optimize the observational efficiency. We generally keep the positions of the stars on the chip within a box of a few pixels of side to improve the photometric precision of our TRAPPIST time-series, thanks to a `software  guiding' system deriving regularly astrometric solutions on the science images and sending pointing corrections to the mount if needed. It could unfortunately not be used for WASP-43,  because the star lies in a sky area that is not covered by the used astrometric catalogue (GSC1.1). This translated 
 into slow drifts of the stars on the chip, the underlying cause being the imperfection of the telescope polar alignment. 
 The amplitudes of those  drifts on the total duration of the runs were ranging between 15 and 85 pixels in the right 
 ascension direction and between 5 and 75 pixels in the declination direction. Table 1 presents the logs of the observations. 
 The first of these 20 transits was presented in H11. 

\begin{table*}
\begin{center}
\scriptsize{
\begin{tabular}{ccccccccccc}
\hline \noalign {\smallskip} 
Date & Instrument & Filter & $N_p$ & Epoch & Baseline &  $\sigma$ & $\sigma_{120s}$ &  $\beta_{w}$ & $\beta_{r}$ & CF \\
         &                  &           &   &                             & function  &  [\%]    &   [\%]                  &                            &                       &        \\     
\hline \noalign {\smallskip} 
06 Dec 2010 & TRAPPIST & {\it I+z} & 393 & 11  & $p(t^2)$ & 0.35, 0.36 & 0.14, 0.17 & 1.26, 1.30 & 1.00, 1.90 & 1.26, 2.47 \\ \noalign {\smallskip} 
09 Dec 2010 & VLT/HAWK-I & NB2090 & 183 & 13.5 & $p(t^2)+p(l^1)+p(xy^1)$ & 0.055, 0.055 &  0.036, 0.036 &  1.04, 1.04 & 1.09, 1.14 & 1.14, 1.18 \\ \noalign {\smallskip} 
15 Dec 2010 & TRAPPIST & {\it I+z} & 442 & 22  & $p(t^2)$ & 0.36, 0.36 & 0.12, 0.14 & 0.97, 0.97 & 1.03, 1.13 & 1.00, 1.10  \\ \noalign {\smallskip} 
19 Dec 2010 & TRAPPIST & {\it I+z} & 480 & 27  & $p(t^2)$ & 0.27, 0.28 & 0.11, 0.12 & 1.00, 1.01 & 1.00, 1.18 & 1.00, 1.19 \\ \noalign {\smallskip} 
28 Dec 2010 & TRAPPIST & {\it I+z} & 574 & 38  & $p(t^2)$ & 0.32, 0.32 & 0.12, 0.12 & 1.01, 1.02 & 1.29, 1.50 & 1.30, 1.53  \\ \noalign {\smallskip} 
28 Dec 2010 & {\it Euler}    & Gunn-$r'$ & 111 & 38  & $p(t^2)+p(xy^2)$ & 0.10, 0.10 & 0.10, 0.10 & 1.54, 1.64 & 1.04, 1.15 & 1.61, 1.90  \\ \noalign {\smallskip} 
30 Dec 2010 & TRAPPIST  & $z'$ & 332 & 40.5 & $p(t^2)$ & 0.30, 0.30 & 0.14, 0.14 & 1.04, 1.04 & 1.00, 1.00 & 1.04, 1.04 \\ \noalign {\smallskip} 
01 Jan 2011  & TRAPPIST  & {\it I+z} & 407 & 43  & $p(t^2)+p(xy^2)$ & 0.27, 0.27 & 0.11, 0.12 & 1.00, 1.01 & 1.00, 1.00 & 1.00, 1.01  \\ \noalign {\smallskip} 
06 Jan 2011 & TRAPPIST & {\it I+z} & 273 & 49  & $p(t^2)$ & 0.24, 0.24 & 0.11, 0.11 & 1.07, 1.09 & 1.00, 1.38 & 1.07, 1.50\\ \noalign {\smallskip} 
09 Jan 2011 & VLT/HAWK-I & NB1190 &  115 & 51.5  & $p(t^2)+ p(b^1)+ p(xy^1)$ & 0.087, 0.087 & 0.047, 0.048 & 2.22, 2.22 & 1.00, 1.00 & 2.22, 2.22 \\ \noalign {\smallskip} 
14 Jan 2011 & TRAPPIST  & {\it I+z} & 237 &  59  & $p(t^2)$ & 0.17, 0.18 &  0.09, 0.10 & 0.91, 0.93 & 1.04, 1.34 & 0.95, 1.24 \\ \noalign {\smallskip} 
19 Jan 2011 & TRAPPIST  & {\it I+z} & 279  & 65  & $p(t^2)$ & 0.18, 0.19 &  0.09, 0.10 & 0.91, 0.95 & 1.00, 2.16 & 0.91, 2.05 \\ \noalign {\smallskip} 
23 Jan 2011 & TRAPPIST  & {\it I+z} & 244  & 70  & $p(t^2)+p(xy^2)$ & 0.20, 0.20 &  0.08, 0.09 & 1.02, 1.02 & 1.00, 1.00 & 1.02, 1.02 \\ \noalign {\smallskip} 
28 Jan 2011 & {\it Euler}    & Gunn-$r'$ & 114 & 76  & $p(t^2)+p(xy^1)$ &  0.13, 0.13 & 0.13, 0.13  &  1.36, 1.42 & 1.17, 1.24 & 1.59, 1.76   \\ \noalign {\smallskip} 
06 Feb 2011 & {\it Euler}     & Gunn-$r'$ & 107 & 87  & $p(t^2)$ &  0.13, 0.15 & 0.13, 0.15 &  0.92, 1.07 & 1.10, 1.86 & 1.02, 1.99  \\ \noalign {\smallskip} 
14 Feb 2011 & TRAPPIST & {\it I+z} & 311  & 97  & $p(t^2)$ & 0.21, 0.22 &  0.11, 0.12 & 1.05, 1.08 & 1.00, 1.42 & 1.05, 1.53 \\ \noalign {\smallskip} 
08 Mar 2011 & TRAPPIST  & {\it I+z} & 285  & 124 & $p(t^2)$ & 0.21, 0.22 &  0.11, 0.13 & 1.11, 1.15 & 1.00, 2.00 & 1.11, 2.30 \\ \noalign {\smallskip} 
10 Mar 2011 & TRAPPIST  & $z'$  & 216 & 126.5 & $p(t^2)$ & 0.22, 0.22 &  0.17, 0.17 & 1.09, 1.09 & 1.00, 1.00 & 1.09, 1.09 \\ \noalign {\smallskip} 
21 Mar 2011 & TRAPPIST  & {\it I+z} & 322  & 140 & $p(t^2)$ & 0.20, 0.21 &  0.11, 0.13 & 0.83, 0.88 & 1.20, 2.04 & 1.00, 1.80 \\ \noalign {\smallskip} 
22 Mar 2011 & TRAPPIST  & {\it I+z} & 237 & 141 & $p(t^2)$ & 0.21, 0.22 &  0.10, 0.11 & 1.10, 1.14 & 1.00, 1.67 & 1.11, 1.91  \\ \noalign {\smallskip} 
28 Mar 2011 & TRAPPIST  & $z'$ & 195  & 148.5 & $p(t^2)$ & 0.28, 0.28 &  0.18, 0.18 & 0.81, 0.81 & 1.03, 1.04 & 0.84, 0.84  \\ \noalign {\smallskip}
31 Mar 2011 & TRAPPIST  & {\it I+z} & 238 & 152 & $p(t^2)$ & 0.24, 0.26 &  0.12, 0.17 & 1.04, 1.15 & 1.00, 2.28 & 1.04, 2.63 \\ \noalign {\smallskip}
02 Apr 2011 & TRAPPIST  & $z'$  & 160 & 154.5 & $p(t^2)+p(b^1)$ & 0.18, 0.18 &  0.11, 0.11 & 0.93, 0.93 & 1.27, 1.30 & 1.19, 1.21 \\ \noalign {\smallskip} 
13 Apr 2011 & TRAPPIST  & {\it I+z} & 293  & 168 & $p(t^2)+p(xy^2)$ & 0.23, 0.23 &  0.12, 0.13 & 1.00, 1.01 & 1.00, 1.00 & 1.00, 1.01  \\ \noalign {\smallskip} 
17 Apr 2011 & TRAPPIST  & {\it I+z} & 289  &  173 & $p(t^2)$ & 0.26, 0.27 & 0.13, 0.15 & 1.05, 1.09 & 1.00, 1.26 & 1.05, 1.36   \\ \noalign {\smallskip} 
30 Apr 2011 & TRAPPIST & {\it I+z} & 327  &  189 & $p(t^2)$ & 0.32, 0.33 & 0.16, 0.17 & 1.31, 1.33 & 1.07, 1.30 & 1.40, 1.73    \\ \noalign {\smallskip} 
09 May 2011 & TRAPPIST  & {\it I+z} & 280 &  200 & $p(t^2)$ & 0.19, 0.20 & 0.10, 0.12 & 0.98, 1.05  & 1.00, 1.77 & 0.98, 1.85   \\ \noalign {\smallskip} 
11 May 2011 & TRAPPIST  & $z'$  & 214 &  202.5 & $p(t^2)$ & 0.21, 0.21 & 0.14, 0.14 & 1.03, 1.03 & 1.23, 1.29 & 1.27, 1.33  \\ \noalign {\smallskip} 
18 May 2011 & TRAPPIST  & {\it I+z} & 250 &  211 & $p(t^2)$ & 0.18, 0.18 & 0.10, 0.11 & 0.93, 0.96 & 1.10, 1.48 & 1.02, 1.42  \\ \noalign {\smallskip} 
13 Jun 2011 & TRAPPIST  & {\it I+z} & 605   &  243 & $p(t^2)$ & 0.33, 0.34 & 0.13, 0.14 & 0.97, 0.99 & 1.25, 1.87 & 1.22, 1.84 \\ \noalign {\smallskip} 
\hline
\end{tabular}}
\caption{WASP-43\,b photometric eclipse time-series used in this work. For each light curve, this table shows the date of 
acquisition, the used instrument and filter, the number of data points, the epoch based on the transit 
ephemeris presented in H11, the selected baseline function (see Sec. 3.1), the standard deviation of the best-fit residuals (unbinned 
and binned per intervals of 2 min), and the deduced values for $\beta_{w}$, $\beta_{r}$, and $CF = \beta_r \times \beta_w$ (see Sec. 3.1). 
For the baseline function, $p(\epsilon^N)$ denotes, respectively, a $N$-order polynomial function of time ($\epsilon=t$), the logarithm of 
time ($\epsilon=l$), $x$ and $y$ positions ($\epsilon=xy$), and background ($\epsilon=b$).  For the last five columns, the first and 
second value correspond, respectively, to the individual analysis of the light curve and to the global analysis of all data.  }
\end{center}
\label{wasp43-phot}
\end{table*}

After a standard pre-reduction (bias, dark, flatfield correction), the stellar fluxes were extracted from the 
images using the {\tt IRAF/DAOPHOT}\footnote{{\tt IRAF} is distributed by the National Optical Astronomy 
Observatory, which is operated by the Association of Universities for Research in Astronomy, Inc., under 
cooperative agreement with the National Science Foundation.} aperture photometry software (Stetson, 1987). 
For each transit, several sets of reduction parameters were tested, and we kept the one giving the most precise 
photometry for the stars of similar brightness as WASP-43. After a careful selection of reference stars, differential 
photometry was then obtained. The resulting light curves are shown in Fig. 1 and 2.

\begin{figure*}
\label{fig:1}
\centering                     
\includegraphics[width=18cm]{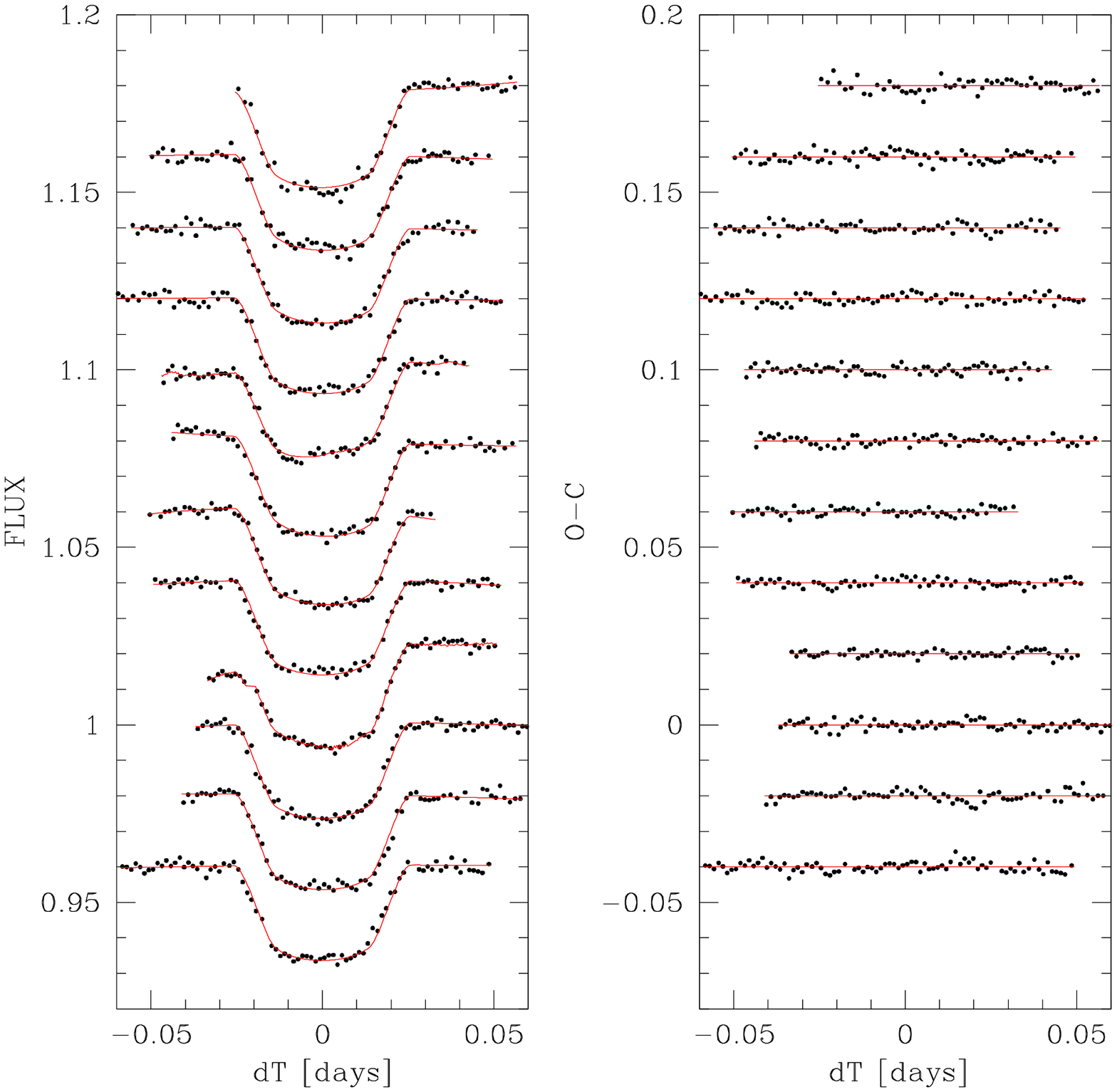}
\caption{$Left:$ WASP-43\,b transit photometry (12 first TRAPPIST transits) used in this work,
 binned per 2 min, period-folded on the best-fit transit ephemeris deduced from our global MCMC analysis 
 (see Sec.~3.3), and shifted along the $y$-axis for clarity. The best-fit baseline+transit models are superimposed on the light curves. The fifth and ninth models (from the top) show some wiggles because of their position-dependent terms. $Right$: best-fit residuals  for each light curve  binned per interval of 2 min.  }
\end{figure*} 

\begin{figure*}
\label{fig:2}
\centering                     
\includegraphics[width=18cm]{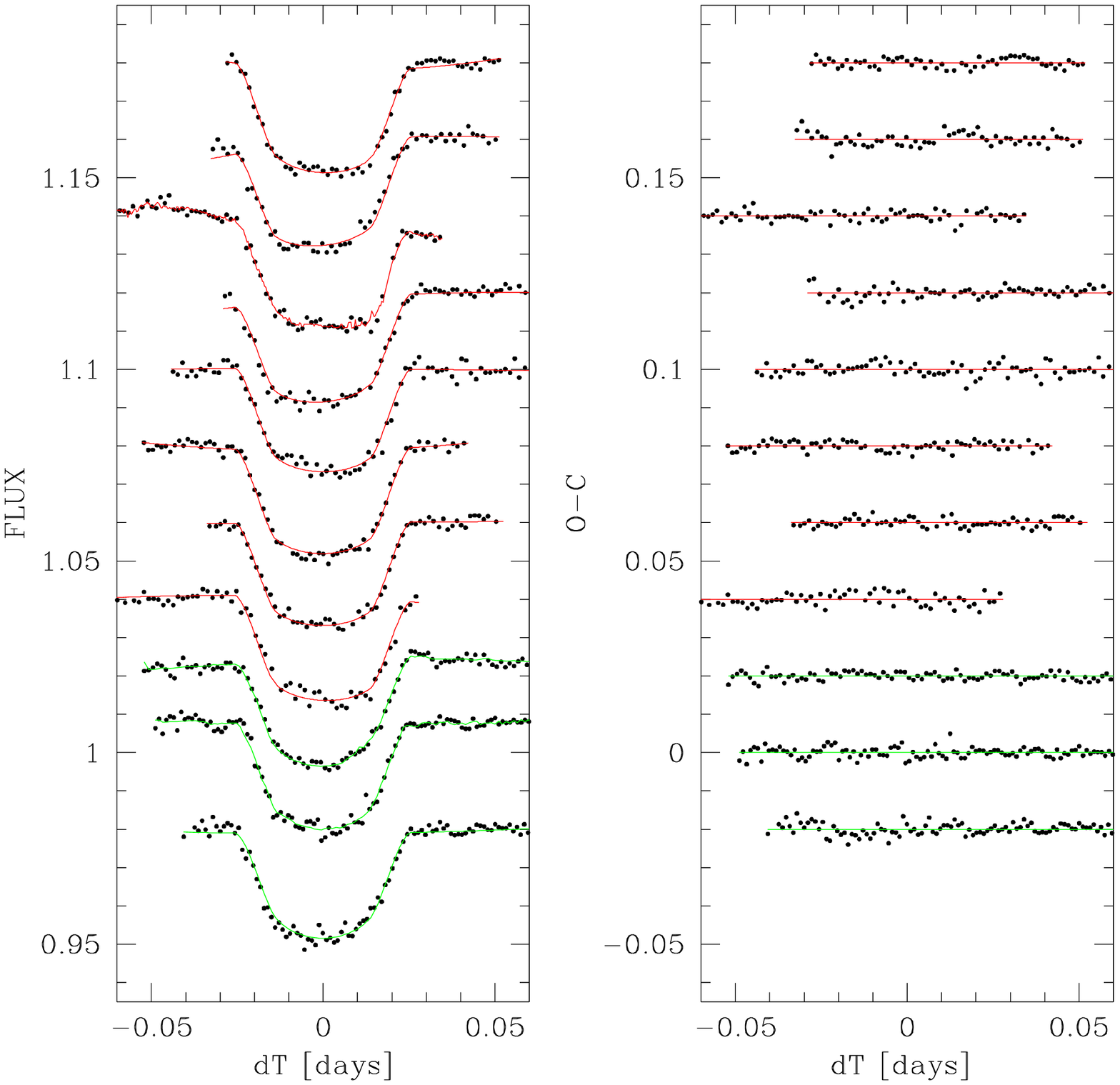}
\caption{$Left:$ WASP-43\,b transit photometry (8 last TRAPPIST transits + the 3 {\it Euler} transits) used in this 
work, binned per 2 min, period-folded on the best-fit transit ephemeris deduced from our global MCMC analysis 
(see Sec.~3.3), and shifted along the  $y$-axis for clarity. The best-fit baseline+transit models are superimposed 
on the light curves  (red = TRAPPIST $I+z$ filter; green = {\it Euler} Gunn-$r'$ filter). The third, ninth and tenth  models (from the top) show some wiggles because of their position-dependent terms. $Right$: best-fit residuals  for each light curve  binned per interval of 2 min.  }
\end{figure*} 
 
 \subsection{{\it Euler} Gunn-$r'$ filter transit photometry}
 
 Three transits of WASP-43\,b were observed in the Gunn-$r'$ filter ($\lambda_{eff} = 620.4 \pm 0.5$ nm) 
 with the EulerCAM CCD camera at the 1.2-m {\it Euler} Swiss telescope, also located at ESO La Silla Observatory. 
 EulerCAM is a nitrogen-cooled 4k $\times$ 4k CCD camera with a 15' $\times$ 15' field of view (pixel scale=0.23"). 
 Here too, the telescope was slightly defocused to optimize the photometric precision. The mean exposure time
 was 85s. The stars were kept approximately on the same pixels, thanks to a `software guiding' system similar to 
 TRAPPIST's but using the UCAC3 catalogue. The calibration and photometric  reduction procedures were similar to the ones 
 performed on the TRAPPIST data. The logs of these {\it Euler} observations are shown in Table 1, while the 
 resulting light curves are visible in Fig.~2. Notice that the first of these {\it Euler} transits was presented in H11. 

\subsection{TRAPPIST $z'$ filter occcultation photometry}

Five occultations of WASP-43\,b were observed with TRAPPIST in a Sloan $z$' filter ($\lambda_{eff} = 915.9 \pm 0.5$ nm). 
Their logs are presented in Table 1. The mean exposure time was 40s. The calibration and photometric reduction of these occultation data were similar to the ones of the transits. Fig.~3 shows the resulting light curves with their best-fit models.

\begin{figure*}
\label{fig:3}
\centering                     
\includegraphics[width=18cm]{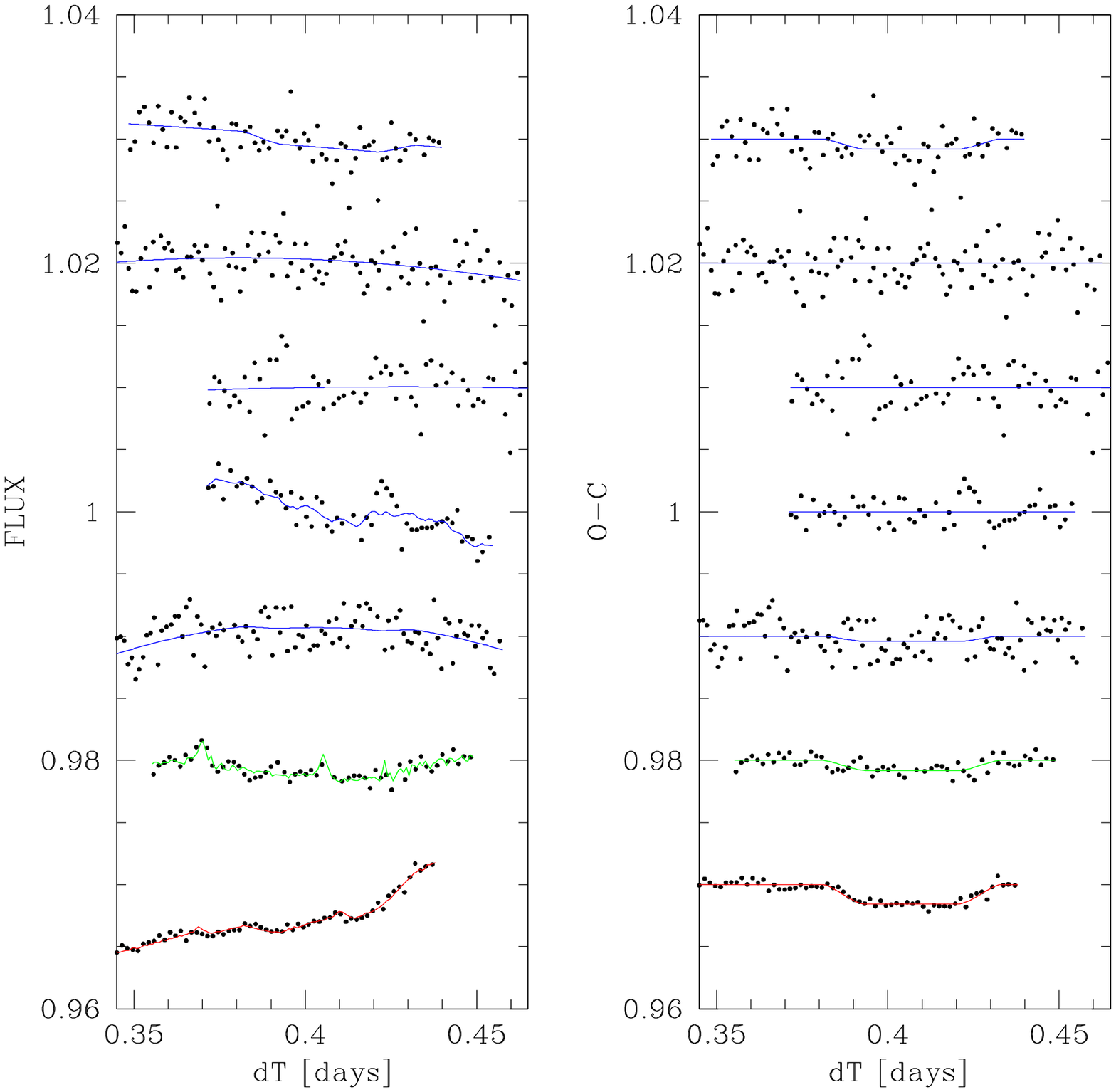}
\caption{$Left:$ WASP-43\,b occultation photometry, binned per intervals of 2 min, period-folded on the 
best-fit transit ephemeris deduced from our global MCMC analysis (see Sec.~3.3), and shifted
along the $y$-axis for clarity. The best-fit baseline+occultation models are superimposed on 
the light curves (blue = TRAPPIST $z'$-filter; green = VLT/HAWK-I NB1190 filter; red = VLT/HAWK-I 
NB2090 filter). $Right$: same light curves divided by their best-fit baseline models. The corresponding best-fit 
occultation models are superimposed. }
\end{figure*}

\subsection{VLT/HAWK-I 1.19 and 2.09 $\mu$m occultation photometry}

We observed two occultations of WASP-43\,b with the cryogenic near-IR imager HAWK-I at the ESO 
Very Large Telescope in our program 086.C-0222. HAWK-I is composed of four Hawaii 2RG 2048$\times$
2048 pixels detectors  (pixel scale = 0.106"), its total field of view on the sky being 7.5'$\times$7.5'.
We choose to observe the occultations within the narrow band filters NB2090 ($\lambda$ = 2.095 $\mu$m, width 
= 0.020  $\mu$m) and NB1190 ($\lambda$ = 1.186 $\mu$m, width = 0.020 $\mu$m), respectively. The 
small width of these cosmological filters minimizes the effect of differential extinction.  Furthermore, they 
avoid the largest absorption and emission bands that are present in $J$ and $K$ bands, reducing thus 
significantly the correlated photometric noise caused by the complex spatial and temporal variations of 
the background due to the variability of the atmosphere. As this correlated noise is  the main precision 
limit for ground-based near-IR time-series photometry, the use of these two  narrow-band filters optimizes 
the photometric quality of the resulting light curves. This is especially important in the context of the 
challenging measurement of the emission of exoplanets. 

The observation of the first occultation (NB2090 filter) took place on  2010 Dec 9 from 5h37 to 9h07 UT. 
Atmospheric conditions were very good, with a stable seeing and extinction. Airmass decreased from 2.1
 to 1.05 during the run. Each of the 185 exposures was composed of 17 integrations of 1.7s each (the 
 minimum integration time allowed for HAWK-I). We choose to do not apply a jitter pattern. Indeed, the 
 background contribution in the photometric aperture is small enough to ensure that the low-frequency 
 variability of the background cannot cause correlated noises with an amplitude larger than a few dozens of 
 ppm in our light curves, making the removal of a background image unnecessary. Furthermore, staying on 
 the same pixels during the whole run allows minimizing the effects of interpixel sensitivity inhomogeneity (i.e. 
 the imperfect flat field). The analysis of HAWK-I calibration frames showed us that the detector is nearly linear 
 up to 10-12 kADU. The peak of the target image was above 10 kADU in the first images, so a slight defocus 
 was applied to keep it below this level during the rest of the run. The mean full-width at half maximum of the 
 stellar point-spread function (PSF) was 7.3 pixels = 0.77", its standard deviation for 
 the whole run being 0.57 pixels = 0.06".  The pointing was carefully selected to avoid cosmetic defects on 
 WASP-43 or on the comparison stars. 
  
The second occultation (NB1190 filter) was observed on 2011 Jan 9 from 4h57 to 9h27 UT.  The seeing 
varied strongly during the all run (mean value in our images =0.76", with a standard deviation = 0.19", 
minimum = 0.51", maximum = 1.22") while the extinction was stable during the first part of the run and 
slightly variable during the second part. No defocus was applied, the peak of the target being in the linear 
part of the detector dynamic in all images. Airmass decreased from 1.37 to 1.03, then increased to 1.13 
during the run. Here too, no jitter pattern was applied. Each of the 241 exposures was composed of 17
 integrations of 1.7s. 

After a standard calibration of the images (dark-subtraction, flatfield correction), a cosmetic correction 
was applied. This correction was done independently for each image and based on an automatic detection 
of the stars followed by a detection of outlier pixels. This latter was based on a comparison of the value of 
each pixel to the median value of the eight adjacent pixels. For a pixel within a stellar aperture, 
a detection threshold of 50-$\sigma$ was used, while it was 5-$\sigma$ for the background pixels. Outlier pixels 
had their value replaced by the median value of the adjacent pixels. At this stage, aperture photometry was 
performed with {\tt DAOPHOT} for the target and comparison stars, and differential photometry was obtained 
for WASP-43. Table 1 presents the logs of our HAWK-I observations, while Fig.~3 shows the resulting light 
curves with their best-fit models.

\subsection{{\it Euler}/{\tt CORALIE} spectra and radial velocities}

We gathered eight new spectroscopic measurements of WASP-43  with the {\tt CORALIE} spectrograph 
mounted on {\it Euler}. The spectroscopic measurements were performed between 2011 Feb 02 and March 12,
 the integration time being 30 min for all of them. RVs were computed from the calibrated spectra by weighted 
 cross-correlation (Baranne et al. 1996) with a numerical spectral template. They are shown in Table 2. We  
 analyzed these new RVs globally with the RVs presented in H11 and with our eclipse photometry (see Sec. 3.3). 

\begin{table}
\begin{center}
\begin{tabular}{cccc}
\hline
Time & RV & $\sigma_{RV}$ & BS\\ 
 ($BJD_{TDB}$-2,450,000) & (km~s$^{-1}$) & (m~s$^{-1}$) & (km~s$^{-1}$)\\ \hline \noalign {\smallskip} 
 5594.848273 & -3.921 &  21 & 0.021  \\ \noalign {\smallskip} 
 5604.730390 & -4.149  &  19 & 0.052  \\ \noalign {\smallskip} 
 5605.677373 &  -3.862 & 15 & 0.044   \\ \noalign {\smallskip} 
 5626.663879 &  -4.123 &  18 & 0.035   \\ \noalign {\smallskip} 
 5627.670365 & -3.759 &  21 & --0.039    \\ \noalign {\smallskip} 
 5628.713064 & -3.059 & 18 & 0.061  \\ \noalign {\smallskip} 
 5629.686781 & -3.388 &  17 & --0.009   \\ \noalign {\smallskip} 
 5632.715791 & -3.126 & 22 &  0.115  \\ \noalign {\smallskip} 
\hline
\end{tabular}
\caption{{\tt CORALIE} radial-velocity measurements for WASP-43 (BS = bisector spans).}
\end{center}
\label{wasp43-rvs}
\end{table}

\section{Data analysis}

Our analysis of the WASP-43 data was divided in two steps.  In a first step, we performed  
individual analyses of the 30 eclipse light curves,  determining independently for each light curves 
the corresponding eclipse and physical parameters. The aim of this first step was  searching for 
potential variability among the eclipse parameters (Sec. 3.2).  We then performed a global  analysis
 of the whole data set, including the RVs, with the aim to obtain the strongest constraints on the system 
 parameters (Sec. 3.3). 
 
\subsection{Method and models}

Our data analysis was based on the most recent version of our adaptive Markov Chain 
Monte-Carlo (MCMC) algorithm (see Gillon et al. 2010 and references therein). To summarize, 
MCMC is a Bayesian stochastic simulation algorithm designed to deduce the posterior Probability 
Distribution Functions (PDFs) for the parameters of a given model (e.g. Gregory 2005, Carlin \& 
Louis 2008). Our implementation of the algorithm assumes as model for the photometric time-series 
the eclipse model of Mandel \& Agol (2002) multiplied by a baseline model aiming to represent the other 
astrophysical and instrumental  mechanisms able to produce photometric variations. For the RVs,
 the model is based on Keplerian orbits added to a model for the stellar and instrumental variability.
Our global model can include any number of planets, transiting or not. For the RVs obtained during a 
transit, a model of the Rossiter-McLaughlin effect is also available (Gim\'enez, 2006). Comparison 
between two models can be performed based on their Bayes factor, this latter being the product of their 
prior probability ratio multiplied by their marginal likelihood ratio. The marginal likelihood ratio of two 
given models is estimated from the difference of their Bayesian Information Criteria (BIC; Schwarz 1978) 
which are given by the formula:
 
 \begin{equation}
 BIC = \chi^2 + k\log(N)
 \end{equation} where $k$ is the number of free parameters of the model, $N$ is the number of 
 data points, and $\chi^2$ is the smallest chi-square found in the Markov chains. From the BIC
 derived for two models, the corresponding marginal likelihood ratio is given by $e^{-\Delta BIC/2}$.

For each planet, the main parameters that can be randomly perturbed at each step of the Markov 
chains (called {\it jump parameters}) are \begin{itemize}
\item the planet/star area ratio $dF = (R_p /R_\star )^2$, $R_p$ and $R_\star$ being respectively 
the radius of the planet and the star; 
\item the occultation depth(s) (one per filter) $dF_{occ}$;
\item the parameter $b' = a \cos{i_p}/R_\star$ which is the transit impact parameter in case of a 
circular orbit, $a$ and $i_p$ being respectively the semi-major axis and inclination of the orbit;
\item the orbital period $P$; 
\item the time of minimum light $T_0$ (inferior conjunction);
\item the two parameters $\sqrt{e} \cos{\omega}$ and $\sqrt{e} \sin{\omega}$, $e$ being the 
orbital eccentricity and $\omega$ being the argument of periastron;
\item the transit width (from first to last contact) $W$; 
\item the parameter $K_2 = K  \sqrt{1-e^2}   \textrm{ }  P^{1/3}$, $K$ being the RV orbital 
semi-amplitude;
\item the parameters $\sqrt{v \sin{I_\star}} \cos{\beta}$ and $\sqrt{v \sin{I_\star}} \sin{\beta}$, 
 $v \sin{I_\star}$ and $\beta$ being respectively the projected rotational velocity of the star and the 
 projected angle between the stellar spin axis and the planet's orbital axis. 
 \end{itemize}
 Uniform or normal prior PDFs can  be assumed for the jump and physical parameters of the system. 
 Negative values are not allowed for $dF$, $dF_{occ}$, $b'$, $P$, $T_0$, $W$ and $K_2$.

Two limb-darkening laws are implemented in our code, quadratic (two parameters) and non-linear 
(four parameters). For each photometric filter, values and error bars for the limb-darkening coefficients 
are interpolated in Claret \& Bloemen's tables (2011) at the beginning of the analysis, basing on input 
values and error bars for the stellar effective temperature $T_\mathrm{eff}$, metallicity [Fe/H] and gravity 
$\log{g}$. For the quadratic law, the two coefficients $u_1$ and $u_2$ are allowed to float in the MCMC, 
using as jump parameters not these coefficients themselves but the combinations $c_1 = 2  \times u_1 
+ u_2$  and $c_2 = u_1 - 2 \times u_2$ to minimize the correlation of the obtained uncertainties 
(Holman et al. 2006). In this case, the theoretical values and error bars for $u_1$ and $u_2$ deduced 
from Claret's tables can be used in normal prior PDFs. In all our analyses, we assumed a quadratic 
law and  let $u_1$ and $u_2$ float under the control of the normal prior PDFs deduced from Claret \& 
Bloemen's tables. For the non-standard $I+z$ filter, the modes of the normal PDFs for $u_1$ and $u_2$ 
were taken as the averages of the values interpolated from Claret's tables for the standard filters $Ic$ and
$z'$, while the errors were computed as the quadratic sums of the errors for these two filters.
The prior PDFs deduced for WASP-43 are  shown in Table 3. They were computed 
for $T_\mathrm{eff} = 4400 \pm 200$K, $\log{g} = 4.5 \pm 0.2$ and [Fe/H] = $-0.05 \pm 0.17$ (H11).

\begin{table}
\begin{center}
\begin{tabular}{ccc}
\hline
Filter & $u_1$ & $u_2$ \\
\hline \noalign {\smallskip} 
 {\it I+z} & $N(0.440,0.035^2)$ & $N(0.180,0.025^2)$ \\ \noalign {\smallskip}
Gunn-$r'$ & $N(0.625,0.015^2)$ & $N(0.115,0.010^2)$ \\ \noalign {\smallskip}
\hline
\end{tabular}
\caption{Prior PDF used in this work for the quadratic limb-darkening coefficients.}
\end{center}
\label{wasp43-ld}
\end{table} 

At the first step of the MCMC, the timings of the measurements are passed to the $BJD_{TDB}$ 
time standard, following the recommendation of Eastman et al. (2010) that outlined that the commonly-used
$BJD_{UTC}$  time standard is not practical for high-precision timing monitoring as it drifts with the addition
 of one leap second roughly each year.

At each step of the Markov Chains, the stellar density is  derived from Kepler's third law and 
the jump parameters $dF$, $b'$, $W$, $\sqrt{e} \cos{\omega}$ and $\sqrt{e} \sin{\omega}$ 
(Seager \& Mallen-Ornelas 2003, Winn 2010). Using as input values the resulting stellar density and values for
$T_{ef f}$ and  [Fe/H] drawn from their prior distributions, a modified version of the stellar 
mass calibration law deduced by Torres et al. (2010) from well-constrained detached binary systems 
(see Gillon et al. 2011b for details) is used to derive the stellar mass. The stellar radius is then 
derived from the stellar density and mass. At this stage, the physical parameters of the planet (mass, 
radius, semi-major axis) are  deduced from the jump parameters and stellar mass and radius.
Alternatively, a value and error for the stellar mass can be imposed at the start of the MCMC analysis, 
In this case, a stellar mass  value is drawn from the corresponding normal distribution at each step of the 
Markov Chains, allowing the code  to deduce the other physical parameters. We preferred here to use
this second option, as WASP-43 was potentially lying at the lower edge of the mass range for which 
the calibration law of Torres et al. is valid, $\sim0.6 M_\odot$. 

If measurements for the rotational period of the star and  for its projected rotational velocity are available, 
they can be used in addition to the stellar radius values deduced at each step of the MCMC to derive a  posterior 
PDF for the inclination of the star (Watson et al. 2010, Gillon et al. 2011b). We did not use this option here despite
that the rotational period of the star was determined  from WASP photometry to be 15.6 $\pm$ 0.4 days (H11),
because the $V\sin{I_\ast}$ measurement presented by H11 ($4.0 \pm 0.4$ \kms) was presented by these
authors as probably affected by a systematic error due to additional broadening of the lines.

Several chains of 100,000 steps were performed for each analysis, their convergences being checked using 
the statistical test of Gelman and Rubin (1992). After election of the best model for a given light curve, a preliminary 
MCMC analysis was performed to estimate the need to rescale the photometric errors. The standard deviation of 
the residuals was compared to the mean photometric errors, and the resulting ratios $\beta_w$ were stored.  
$\beta_w$ represents the under-  or overestimation of the white noise of each measurement. On its side, the red 
noise present in the light curve (i.e. the inability of our model to represent perfectly the data) was taken into 
account as described by Gillon et al. (2010), i.e. a scaling factor $\beta_r$ was determined from the standard 
deviations of the binned and unbinned residuals for different binning intervals ranging from 5 to 120 minutes, the 
largest values being kept as $\beta_r$. At the end, the error bars  were multiplied by the correction factor $CF = 
\beta_r \times \beta_w$.  For the RVs, a  `jitter' noise could be added quadratically to the error bars after the 
election of the best model, to equal the mean error with the standard deviation of the best-fit model residuals. 
In this case, it was unnecessary.

Our MCMC code can model very complex  trends for the photometric and  RV time-series, 
with up to 46 parameters for each light curve and 17 parameters for each RV time-series. Our strategy here was first to fit 
a simple orbital/eclipse model and to analyze the residuals to assess any correlation with the external 
parameters (PSF width, time, position on the chip, line bisector, etc.), then to use the Bayes factor as indicator 
to find the optimal baseline function for each time-series, i.e. the model minimizing the number of parameters 
and the level of correlated noise in the best-fit residuals. For ground-based photometric time-series, we 
did not use a model simpler than a quadratic polynomial in time, as several effects (color effects, PSF variations, 
drift on the chip, etc) can distort slightly the eclipse shape and can thus lead to systematic errors on the deduced 
transit parameters. This is especially important to include a trend in the baseline model for transit light curve with 
no out-of-transit data before or after the transit, and/or with a small amount of out-of-transit data. Having a rather 
small amount of out-of-transit data is quite common for ground-based transit photometry, as the target star is visible 
under good conditions (low airmass) during a limited duration per night. In such conditions, an eclipse model can 
have enough degrees of  freedom to compensate for a small-amplitude trend in the light curve, leading to an 
excellent fit but also to biased results and overoptimistic error bars. Allowing the MCMC to `twist' slightly the light 
curve with a quadratic polynomial in time compensates, at least partially, for this effect.

For the RVs, our minimal baseline model is a scalar $V_\gamma$ representing the systemic velocity of the star. 
It is worth noticing that most of the baseline parameters are not jump parameters in the MCMC, they are 
deduced by least-square minimization from the residuals at each step of the chains, thanks to their linear nature in 
the baseline functions (Bakos et al. 2009, Gillon et al. 2010). 

\subsection{Individual analysis of the eclipse time-series} 

As mentioned above, we first performed an independent analysis of the transits and occultations aiming to elect 
the optimal model for each light curve and to assess the variability and robustness of the derived parameters. 
For all these analyses, the orbital period and eccentricity were kept respectively to 0.813475 days and zero (H11),
 and the normal distribution $N(0.58,0.05^2)$ $M_\odot$ (H11) was used as prior PDF for the stellar mass. For 
 the transit light curves, the jump parameters were $dF$, $b'$, $W$ and $T_0$. For the occultations, the only 
 jump parameter was the occultation depth $dF_{occ}$, the other system parameters being drawn at each step 
 of the MCMC from normal distributions deduced from the values + errors derived in H11. 
  
 Table 1 presents the baseline function  selected for each light curve, the derived factors $\beta_{w}$, $\beta_{r}$, 
 $CF$, and the standard deviation of the best-fit residuals, unbinned and binned per 120s. These results allow us
  to assess the photometric precision of the used instruments.The TRAPPIST data show mean values for $\beta_{w}$ 
  and $\beta_{r}$ very close to 1, the $I+z$ data having $<\beta_{w}> = 1.03$ and $<\beta_{r}> = 1.05$, while the $z'$
data have  $<\beta_{w}> = 0.98$ and $<\beta_{r}> = 1.11$. This suggests that the photometric errors of each 
measurement are well approximated by a basic error budget (photon, read-out, dark, background, scintillation noises), 
and that the level of correlated noise in the data is small. Furthermore, we notice that most TRAPPIST light curves are 
well modeled by the `minimal model', i.e. the sum of an eclipse model and a quadratic trend in time. Only one TRAPPIST 
transit light curve requires additional terms in $x$ and $y$, while one occultation light curve acquired when the moon 
was close to full requires a linear term in background. The mean photometric errors per 2 min intervals can also be 
considered as very good for a 60cm telescope monitoring a $V=12.4$ star: 0.11\% and 0.15\% in the $I+z$ and $z'$ filters, 
 respectively, which is similar to the mean photometric error of {\it Euler} data, 0.12\%. {\it Euler} data show also small 
 mean values of $<\beta_{w}> = 1.27$ and $<\beta_{r}> = 1.05$. Their modeling requires PSF position terms for 2 out 
 of 3 eclipses, despite the good sampling of the PSF and the active guiding system keeping the stars nearly on the same 
 pixels. This could indicate that the flatfields'  quality is perfectible. 
 
For the first HAWKI light curve, taken in the NB2090 filter, we notice that we have to account for a `ramp' effect (the $\log(t)$ 
term) similar to the well-documented sharp variation of the effective gain of the {\it Spitzer}/IRAC detector at 8 $\mu$m (e.g. 
Knutson et al. 2008), and also for a dependance of the measured flux with the exact position of the PSF center on the chip. 
This position effect could be decreased by spreading the flux on more pixels (defocus), but this would also increase the 
background's contribution to the noise budget, potentially bringing not only more white noise but also some correlation of 
the measured counts with the variability of the local thermal structure. The photometric quality reached by these NB2090 
data ($\beta_{r}$ = 1.09, mean error of 360 ppm per 2 min time interval), can be judged as excellent. For the NB1190 HAWKI 
data, we had to include a dependance in the PSF width in the model. This is not surprising, considering the large variability of 
the seeing during the run. We also notice for these data that our error budget underestimated strongly the noise of the 
measurements ($\beta_{w}=2.22$), suggesting an unaccounted noise source. Still, the reached photometric quality remains 
excellent ($\beta_{r}$= 1, mean error of 470 ppm per 2 min time interval).    
 
 Table 4 presents for each eclipse light curve the values and errors deduced for the jump parameters. Several points can be 
 noticed.\begin{itemize}
 \item The emission of the planet is clearly (10-$\sigma$) detected at 2.09 $\mu$m. At 1.19 $\mu$m, it is  barely detected 
 ($\sim$ 2.3-$\sigma$). It is not detected in any of the $z'$-band light curves.
 \item The transit shape parameters $b'$, $W$, and $dF$, show scatters  slightly  larger than their average error, the 
 corresponding ratio being, respectively, 1.24, 1.26, and 1.54. Furthermore, these parameters show significant correlation, 
 as can be seen in Fig.  4. 
\item Fitting a transit ephemeris by linear regression with the measured transit timings shown in Table 4, we obtained $T(N) = 
2,455,528.868289 (\pm 0.000072) + N \times 0.81347728 (\pm 0.00000060)$ BJD$_{TDB}$. Table 4 shows (last column) the 
resulting transit timing residuals (observed minus computed, O-C). Fig.~5 (upper panel) shows them as a function of the transit 
epochs. The scatter of these timing residuals is 2.1 times larger than the mean error, 18s. No clear pattern is visible in the O-C diagram.
 \end{itemize}
 The apparent variability of the four transit parameters is present in both TRAPPIST and {\it Euler} results. The correlation of the 
 derived transit parameters  is not consistent with actual variations of these parameters and favors biases from instrumental and/or 
 astrophysical origin. This apparent variability of the transit parameters could be explained by the variability of the star itself. Indeed, 
 WASP-43 is a spotted  star (H11), and occulted spots (and faculae)  can alter the observed transit shape and bias the measured transit parameters (e.g. Huber et al. 2011, Berta et al. 2011). We do not detect clear spot signatures in our light curves, but our photometric precision could be not high enough to detect such low-amplitude structures, so we do not reject this hypothesis.
 On their side, unocculted spots should have a negligible impact on the measured  transit depths, considering the low amplitude of the
  rotational photometric modulation detected in WASP data ($6 \pm 1$ mmag). Still, they could alter the slope of the photometric baseline
   and make it more complex. We represent this baseline as an analytical function of several external parameters in our MCMC simulations, 
   but the unavoidable inaccuracy of the chosen baseline model can also bias the derived results. As described in Sec. 2.1, no pointing 
   corrections were applied during the TRAPPIST runs because of a stellar catalogue problem. Even if the resulting drifts did not introduce 
   clear correlations of the measured fluxes with PSF positions, they could have slightly affected the shape of the transits. These considerations 
   reinforce the interest of performing global analysis of extensive data sets (i.e. many eclipses) in order to minimize systematic errors and to
    reach high accuracies on the derived parameters. 
  
\begin{figure}
\label{fig:4}
\centering                     
\includegraphics[width=9cm]{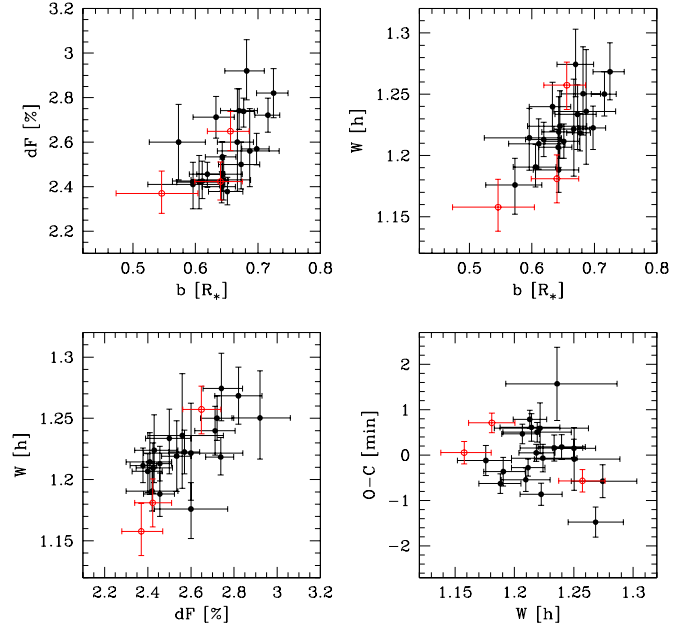}
\caption{Correlation diagrams for the transit parameters deduced from the individual MCMC analysis of the 
23 transit light curves. {\it Top left}: transit depth $vs$ transit impact parameter. {\it Top right}: transit duration 
$vs$ transit impact parameter. {\it Bottom left}: transit duration $vs$ transit depth. {\it Bottom right}: TTV (observed 
minus calculated transit timing)  $vs$ transit duration. The filled black and open red symbols correspond, respectively,
 to the TRAPPIST and {\it Euler} light curves.}
\end{figure}

\begin{figure}
\label{fig:5}
\centering                     
\includegraphics[width=9cm]{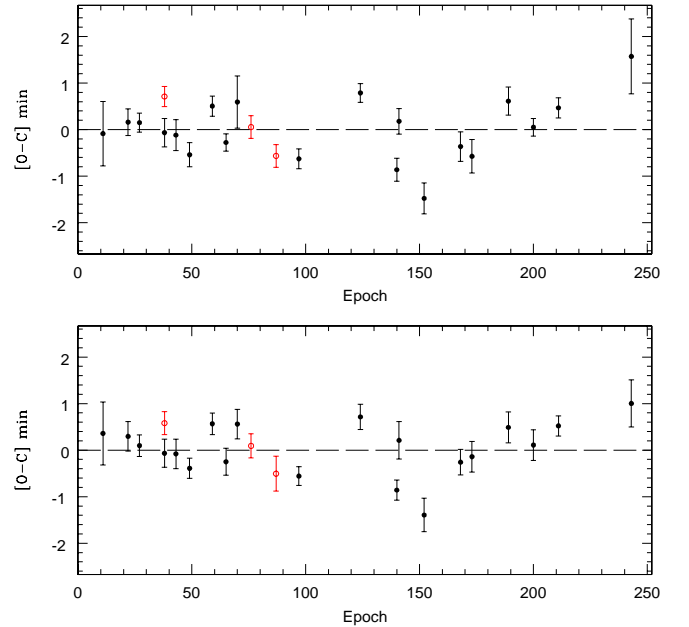}
\caption{$Top$: Observed minus calculated transit timings obtained from the individual analysis of the transit 
light curves as a function of the transit epoch. The filled black and open red symbols correspond, respectively, to
  the TRAPPIST and {\it Euler} light curves. $Bottom$: same but deduced from the global analysis of all
  transits.}
\end{figure}

\begin{table*}
\begin{center}
\begin{tabular}{ccccccccc}
\hline \noalign{\smallskip} 
Epoch & Filter & $dF_{occ}$  & $b$ & $dF$ & $W$ &             $T_0$                             & O-C                          \\
  &   & [\%] &         &  [\%]  &   [h]  & [BJD$_{TDB}$-2450000]    & [min]               \\     \noalign{\smallskip}  
\hline \noalign{\smallskip} 
11 & $I+z$  & & $0.682_{-0.035}^{+0.028}$  & $2.92_{-0.13}^{+0.14}$  & $1.250_{-0.034}^{+0.038}$ & 
$5537.81648_{-0.00048}^{+0.00043}$ & $-0.09 \pm 0.69$ & 
\\ \noalign {\smallskip} 
13.5 & NB2090 & $0.156_{-0.016}^{+0.015}$ 
\\ \noalign {\smallskip} 
22 & $I+z$ & & $0.673_{-0.036}^{+0.030}$ &  $2.50_{-0.11}^{+0.10}$ & $1.234_{-0.029}^{+0.024}$ &  
$5546.76490 \pm 0.00020$ &  $0.16 \pm 0.29$ &  
 \\ \noalign {\smallskip} 
27 & $I+z$  & &  $0.716_{-0.021}^{+0.019}$ &  $2.721 \pm 0.077$ & $1.250 \pm 0.018$ &  
$5550.83228_{-0.00013}^{+0.00014}$ & $0.15 \pm 0.20$  & 
\\ \noalign {\smallskip} 
38 & $I+z$  & & $0.645_{-0.052}^{+0.043}$ &  $2.43_{-0.09}^{+0.11}$ & $1.224_{-0.026}^{+0.029}$ &  
$5559.78038 \pm 0.00021$ & $-0.07 \pm 0.30$ &  
\\ \noalign {\smallskip} 
38 & Gunn-$r'$ & & $0.640_{-0.041}^{+0.035}$ &  $2.423_{-0.083}^{+0.088}$ & $1.181_{-0.020}^{+0.019}$ &  
$5559.78092 \pm 0.00016$ & $0.71 \pm 0.22$ & 
\\ \noalign {\smallskip} 
40.5 &  $z'$ &  $0.090_{-0.053}^{+0.061}$
\\ \noalign {\smallskip} 
43  & $I+z$ &  & $0.573_{-0.047}^{+0.043}$ &  $2.60 \pm 0.17$ & $1.176_{-0.024}^{+0.022}$ & 
 $5563.84773 \pm 0.00023$ & $-0.12 \pm 0.33$ &  
\\ \noalign {\smallskip} 
49  & $I+z$ & & $0.611_{-0.048}^{+0.042}$ &  $2.425_{-0.079}^{+0.090}$ & $1.210_{-0.021}^{+0.020}$ &  
$5568.72830_{-0.00017}^{+0.00018}$ & $-0.54 \pm 0.26$ & 
\\ \noalign {\smallskip}
51.5 & NB1190 & $0.082_{-0.037}^{+0.035}$
\\ \noalign {\smallskip}
59 & $I+z$ &  & $0.643_{-0.027}^{+0.022}$ &  $2.534_{-0.070}^{+0.067}$ & $1.219 \pm 0.029$ &  
$5576.86380 \pm 0.00015$ & $0.51 \pm 0.22$ & 
\\ \noalign {\smallskip} 
65  & $I+z$ &  & $0.651_{-0.030}^{+0.025}$ &  $2.378_{-0.060}^{+0.064}$ & $1.212 \pm 0.014$ &  
$5581.74412_{-0.00013}^{+0.00012}$ & $-0.28 \pm 0.19$  & 
\\ \noalign {\smallskip}
70 & $I+z$ & & $0.667_{-0.032}^{+0.026}$ &  $2.60_{-0.22}^{+0.24}$ & $1.222_{-0.038}^{+0.041}$ &  
$5585.81211_{-0.00038}^{+0.00039}$ & $0.59 \pm 0.56$ & 
\\ \noalign {\smallskip} 
76 & Gunn-$r'$ & & $0.546_{-0.073}^{+0.058}$ &  $2.37_{-0.09}^{+0.10}$ & $1.158_{-0.020}^{+0.023}$ &  
$5590.69260_{-0.00016}^{+0.00017}$ & $0.05 \pm 0.25$ &  
\\ \noalign {\smallskip} 
87 & Gunn-$r'$ &  & $0.656_{-0.037}^{+0.030}$ &  $2.649_{-0.087}^{+0.089}$ & $1.257_{-0.020}^{+0.019}$ & 
 $5599.64042 \pm 0.00017$ & $-0.57 \pm 0.25$ & 
\\ \noalign {\smallskip}
97 & $I+z$ &  &$0.643_{-0.041}^{+0.032}$ &  $2.456_{-0.075}^{+0.068}$ & $1.188_{-0.018}^{+0.017}$ &  
$5607.77515_{-0.00015}^{+0.00014}$ & $-0.63 \pm 0.22$  & 
\\ \noalign {\smallskip}
124 & $I+z$ & &  $0.619_{-0.029}^{+0.028}$ &  $2.456_{-0.059}^{+0.064}$ & $1.213 \pm 0.014$ & 
 $5629.74002_{-0.00013}^{+0.00014}$ &$0.79 \pm 0.20$ & 
\\ \noalign {\smallskip} 
126.5 &  $z'$ & $0.034_{-0.023}^{+0.033}$
\\ \noalign {\smallskip} 
140 & $I+z$ & & $0.698_{-0.024}^{+0.020}$ &  $2.570_{-0.072}^{+0.070}$ & $1.223 \pm 0.018$ & 
 $5642.75451_{-0.00017}^{+0.00016}$ & $-0.86 \pm 0.25$ & 
\\ \noalign {\smallskip} 
141 & $I+z$  & &  $0.633_{-0.037}^{+0.029}$ &  $2.712_{-0.094}^{+0.093}$ & $1.240 \pm 0.020$ &  
$5643.56871_{-0.00018}^{+0.00019}$ &  $0.18 \pm 0.27$ & 
\\ \noalign {\smallskip} 
148.5 &  $z'$ & $0.039_{-0.027}^{+0.042}$
\\ \noalign {\smallskip} 
152 & $I+z$ & & $0.725_{-0.027}^{+0.023}$ &  $2.82 \pm 0.11$ & $1.268_{-0.023}^{+0.024}$ &  
$5652.51581_{-0.00023}^{+0.00021}$ & $-1.48 \pm 0.33$  & 
\\ \noalign {\smallskip}
154.5 &  $z'$ & $0.042_{-0.029}^{+0.050}$
\\ \noalign {\smallskip} 
168  & $I+z$ & & $0.606_{-0.036}^{+0.033}$ &  $2.42 \pm 0.12$ & $1.191_{-0.016}^{+0.018}$ & 
 $5665.53222_{-0.00021}^{+0.00022}$ & $-0.36 \pm 0.32$ & 
\\ \noalign {\smallskip}
173 &  $I+z$ & & $0.670_{-0.030}^{+0.029}$ &  $2.741_{-0.084}^{+0.098}$ & $1.274_{-0.026}^{+0.029}$ & 
 $5669.59946 \pm 0.00025$ & $-0.58 \pm 0.36$ & 
\\ \noalign {\smallskip}
189 &  $I+z$ & & $0.596_{-0.072}^{+0.051}$ &  $2.41_{-0.11}^{+0.10}$ & $1.214_{-0.026}^{+0.024}$ &  
$5682.61592_{-0.00021}^{+0.00020}$ &$0.61 \pm 0.30$ & 
 \\ \noalign {\smallskip}
200 & $I+z$ & & $0.677_{-0.021}^{+0.017}$ &  $2.738_{-0.070}^{+0.060}$ & $1.218_{-0.015}^{+0.016}$ &  
$5691.56378_{-0.00012}^{+0.00013}$ & $0.05 \pm 0.19$  & 
\\ \noalign {\smallskip}
202.5 &  $z'$ & $0.065_{-0.044}^{+0.054}$
\\ \noalign {\smallskip}
211 & $I+z$ & &  $0.642_{-0.031}^{+0.022}$ &  $2.400_{0.073}^{+0.067}$ & $1.207_{-0.017}^{+0.014}$ &  
$5700.51232_{-0.00014}^{+0.00015}$ & $0.47 \pm 0.22$  & 
\\ \noalign {\smallskip}
243  & $I+z$ & & $0.687_{-0.052}^{+0.047}$ &  $2.56_{-0.16}^{+0.19}$ & $1.236_{-0.043}^{+0.050}$ & 
 $5726.54436_{-0.00044}^{+0.00056}$ & $1.57 \pm 0.81$ 
 \\ \noalign {\smallskip}
\hline
\end{tabular}
\caption{Median and 1-$\sigma$ errors of the posterior PDFs  deduced for the jump parameters from the individual analysis 
of the eclipse light curves. For each light curve, this table shows the epoch based on the transit ephemeris presented in H11,
the filter, and the derived values for the occultation depth, impact parameter, transit depth, transit duration, and transit time
of minimum light. The last column shows for the transits the difference (and its error) between the measured timing 
and the one deduced from the best-fitting transit ephemeris computed by linear regression, $T(N) = 2455528.868289 (0.000072) 
+ N\times 0.81347728 (0.00000060)$ BJD$_{TDB}$.  }
\end{center}
\label{wasp43-indi}
\end{table*}

 \subsection{Global analysis of photometry and radial velocities}
 
The global MCMC analysis of our data set was divided in three steps. For each step of the analysis, the MCMC jump parameters 
were $dF$, $W$, $b'$, $T_0$, $P$, $\sqrt{e} \cos{\omega}$, $\sqrt{e} \sin{\omega}$, three $dF_{occ}$ (for the $z'$, NB1190, and 
NB2090 filters), $K_2$, and the limb-darkening coefficients $c_1$ and $c_2$ for the $I+z$ and Gunn-$r'$ filters. No model for the 
Rossiter-McLaughlin was included in the global model, as no RV was obtained at the transit phase. For each light curve, we assumed 
the same baseline model than for the individual analysis. The assumed baseline for the RVs was a simple scalar (systemic 
velocity $V_\gamma$).

In a first step, we performed a chain of 100,000 steps with the aim to redetermine the scaling factors of the photometric errors. 
The deduced values are shown in Table 1. It can be noticed that the mean $<\beta_r>$ for the 20 transits observed by TRAPPIST
 in $I+z$ filter is now 1.52, for 1.05 after the individual analysis of the lightcurves. We notice the same tendency for the {\it Euler} 
 Gunn-$r'$ transits: $<\beta_r>$ goes up from 1.05 to 1.42. Considering the apparent variability of the transit parameters deduced 
 from the individual analysis of the light curves and their correlations, our interpretation of this increase of the $<\beta_r>$  is that 
 a part of the correlated noise (from astrophysical source or not) of a transit light curve can be `swallowed' in the transit+baseline model, leading to a good fit in terms of merit function but also to biased derived parameters. By relying on the assumption that all transits share the same profile, the  global analysis allows to better separate the actual transit signal from the correlated noises of similar frequencies, leading to larger   $<\beta_r>$ values. It is also worth noticing that the  $<\beta_r>$ for the transits, 1.50, is larger than the one for the occultations,  1.12, which is consistent with the crossing of spots by the planet during transit, but can also be explained by the much larger number of transits than occultations. 

In a second step, we used the updated error scaling factors and performed a MCMC of 100,000 steps to derive the marginalized 
PDF for the stellar density $\rho_\star$. This PDF was used as input to interpolate stellar tracks. As in Hebb et al. (2009), we used 
$\rho_\star ^{-1/3}$, mostly for practical reasons. Other inputs were the stellar effective and metallicity derived by H11, $T_\mathrm{eff} 
= 4400 \pm 200$ K and [Fe/H] = -0.05 $\pm$ 0.17 dex. Those parameters were assumed distributed as a Gaussian. For each $\rho_\star ^{-1/3}$ 
value from the MCMC, a random Gaussian value was drawn for stellar parameters. That point was placed among the Geneva stellar 
evolution tracks (Mowlavi et al. 2012) that are available in a grid fine enough to allow an MCMC to wander without computational
systematic effects; a mass and age were interpolated. If the point fell off the tracks, it was discarded.  Because of 
the way stars behave at these small masses, within the ($\rho_\star ^{-1/3}$, $T_\mathrm{eff}$, [Fe/H]) space, the parameter space over 
which the tracks were interpolated covered stellar ages well above a Hubble time. A restriction on the age was therefore added. Only taking ages 
$<$ 12 Gyr, we obtained the distribution as shown in Fig.~6, refining the stellar parameters to $T_\mathrm{eff} = 4520 \pm 120$ K and 
[Fe/H] = -0.01 $\pm$ 0.12 dex. The stellar mass is estimated to $M_\star = 0.717 \pm 0.025 M_\odot$. The output distribution is very close 
to Gaussian as displayed in Fig.~6. In addition the output $\rho_\star ^{-1/3}$ distribution has not been affected. Thus, our strongly improved 
precision on the stellar density combined with a constraint on the age (to be younger than a Hubble time) allows us to reject the main stellar solution 
proposed by H11 ($M_\star = 0.58 \pm 0.05 M_\odot$). Because the allowed ($\rho_\star ^{-1/3}$, $T_\mathrm{eff}$, [Fe/H]) space within the 
tracks still covers quite a large area, it is impossible at the moment to constrain the stellar age. The process is described  in Triaud 2011 (PhD thesis).

\begin{figure}
\label{fig:6}
\centering                     
\includegraphics[width=9cm,angle=0]{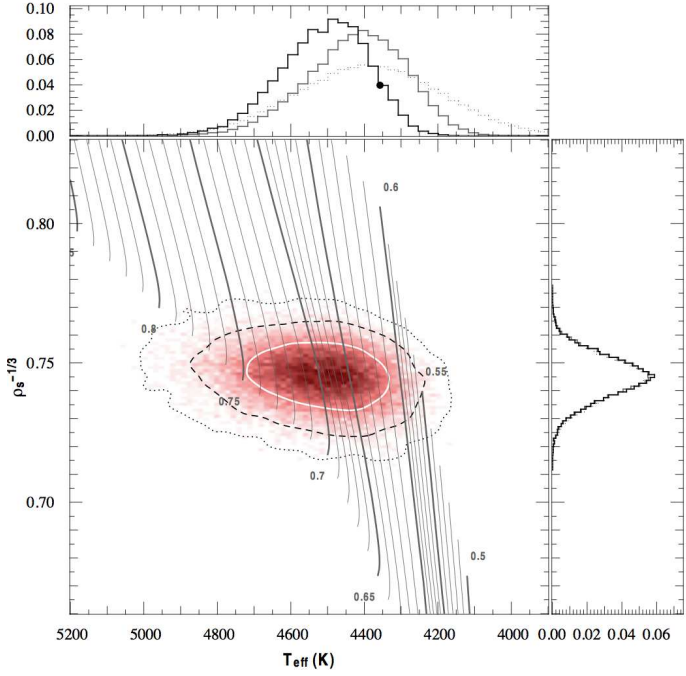}
\caption{{\it Main panel}: modifield Herztsprung-Russell diagram showing the posterior PDF of $\rho^{-1/3}_\star$ as a function of 
$T_\mathrm{eff}$ after interpolating with the Geneva stellar evolution tracks for stars on the main sequence. Masses are indicated and 
correspond to the bold tracks. The solid, dashed and dotted lines correspond to the 1-$\sigma$, 2-$\sigma$ and 3-$\sigma$ contours, 
respectively. {\it Secondary panels}: marginalised PDF of $\rho^{-1/3}_\star$ and $T_\mathrm{eff}$. In both panels, 
there are three histograms: dotted: input distribution; grey: output distribution; black: output distribution for ages $<$ 12 Gyr.}
\end{figure}

We then performed a third MCMC step, using as prior distributions for $M_\star$, T$_\mathrm{eff}$, and [Fe/H], the normal distributions 
matching the results of our stellar modeling, i.e. $N(0.717,0.025^2)$ $M_\odot$, $N(4520,120^2)$ K, and $N(-0.01,0.12^2)$ dex, respectively. 
The best-fit photometry and RV models are shown respectively in Fig.~7 and 8, while the derived parameters and 1-$\sigma$ error bars are shown in Table 5.  

\begin{figure*}
\label{fig:7}
\centering                     
\includegraphics[width=9cm,angle=0]{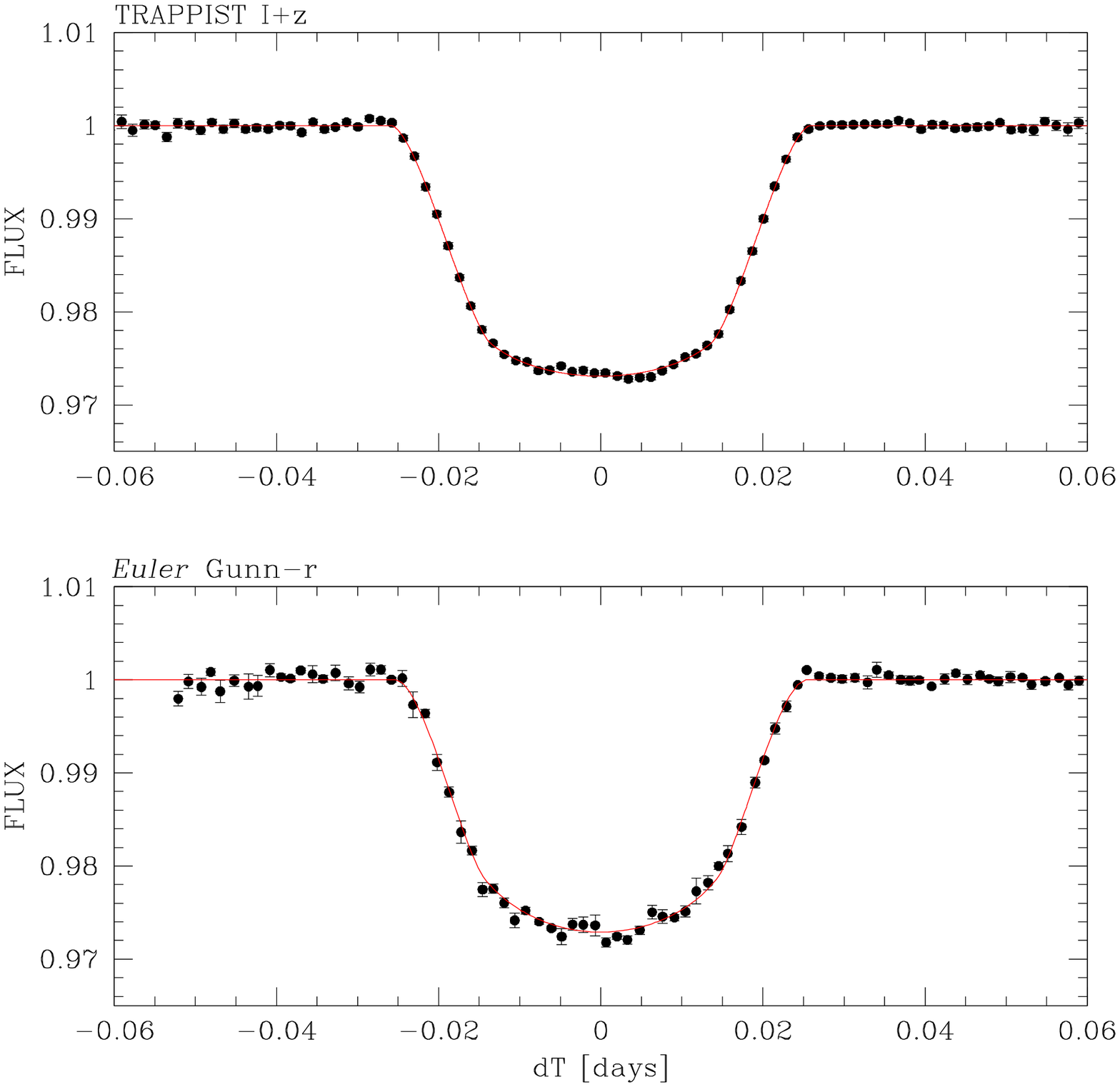}
\includegraphics[width=9cm,angle=0]{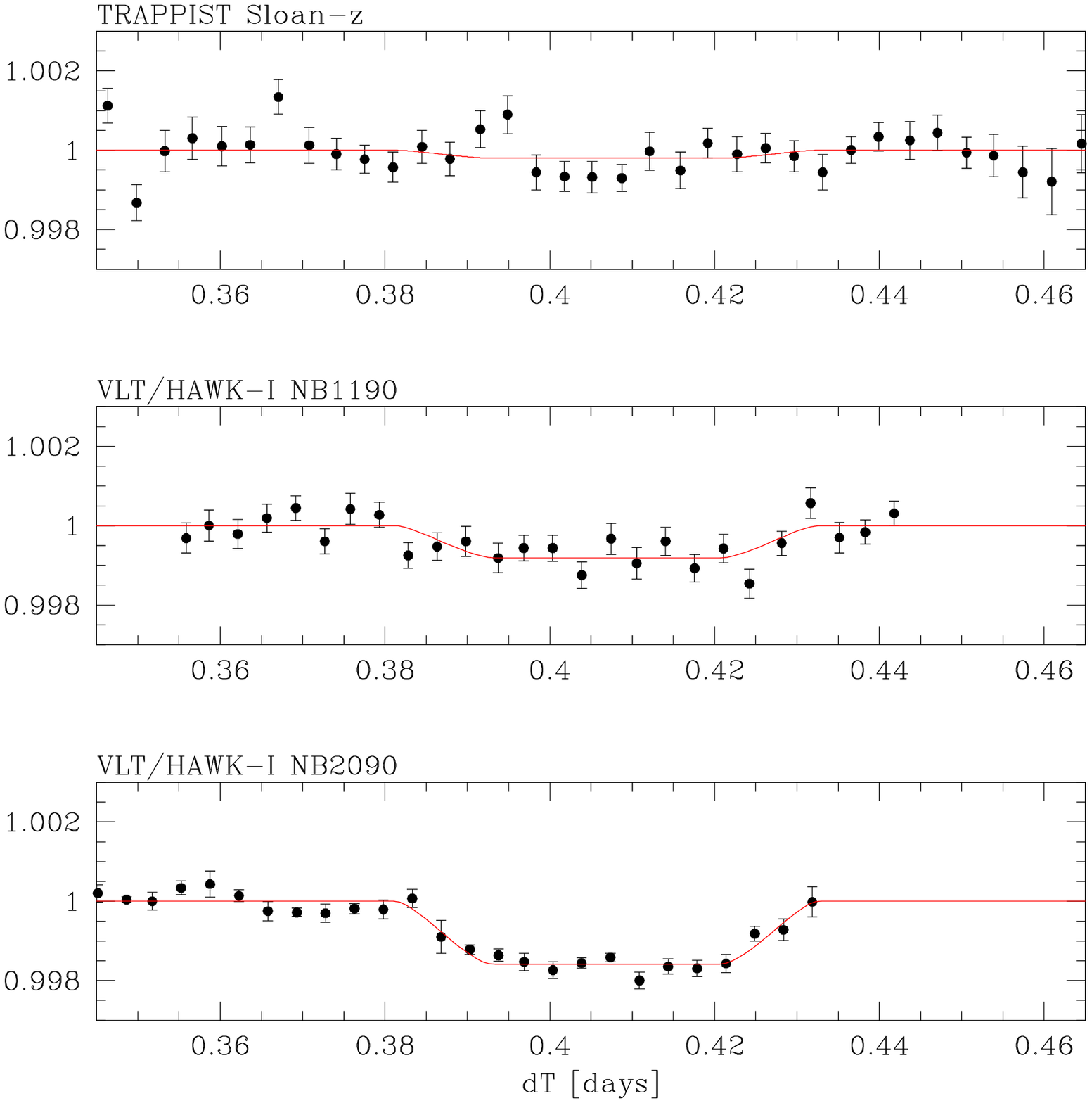}
\caption{{\it Left}: TRAPPIST $I+z$ (top) and {\it Euler} Gunn-$r'$ (bottom) transit photometry period-folded on the best-fit transit 
ephemeris from the global MCMC analysis, corrected for the baseline and binned per 2 min intervals, with the best-fit
transit models over-imposed. {\it Right:} same for the occultation photometry obtained by TRAPPIST in $z'$ filter 
(top), and by VLT/HAWK-I in NB1190 (middle) and NB2090 (bottom) filters, except that the data points are binned per 5 min 
intervals.}
\end{figure*}

\begin{figure}
\label{fig:8}
\centering  
\includegraphics[width=9cm,angle=0]{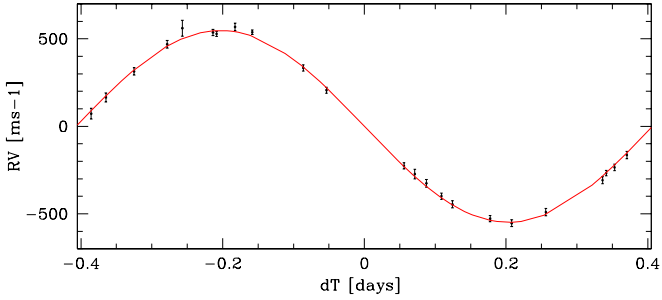}
\caption{ {\it Euler}/{\tt CORALIE} RV measurements period-folded on the best-fit 
transit ephemeris from the global MCMC analysis, corrected for the systemic RV, 
with the best-fit orbital model overimposed.}
\end{figure}

\begin{table}
\begin{center}
\begin{tabular}{cc}
\hline \noalign {\smallskip}
MCMC $Jump$ $parameters$ &  \\ \noalign {\smallskip}
\hline \noalign {\smallskip}
Planet/star area ratio  $ (R_p/R_\star)^2$ [\%]     &  $2.542_{-0.025}^{+0.024}$                      \\ \noalign {\smallskip} 
$b'=a\cos{i_p}/R_\star$ [$R_\star$]                      &  $0.656 \pm 0.010$                                   \\ \noalign {\smallskip} 
Transit width  $W$ [h]                                          &   $1.2089_{-0.0050}^{+0.0055}$          \\ \noalign {\smallskip} 
$T_0-2450000$ [$BJD_{TDB}$]                          &  $5726.54336 \pm 0.00012$                      \\ \noalign {\smallskip}
Orbital period  $ P$ [d]                                         &  $0.81347753 \pm 0.00000071$                \\ \noalign {\smallskip} 
RV $K_2$  [m\,s$^{-1}$\,d$^{1/3}$]                      & $511.5_{-5.0}^{+5.1}$                                \\ \noalign {\smallskip} 
$\sqrt{e}\cos{\omega}$                                         &  $0.020_{-0.023}^{+0.022}$                       \\ \noalign {\smallskip} 
$\sqrt{e}\sin{\omega}$                                          &  $-0.025_{-0.064}^{+0.066}$                      \\ \noalign {\smallskip} 
$dF_{occ,z'}$ [ppm]                                              &  $210_{-130}^{+190}$, $<850$ (3-$\sigma$)  \\ \noalign {\smallskip} 
$dF_{occ,NB1190}$                                              &  $790_{-310}^{+320}$, $<1700$ (3-$\sigma$)   \\ \noalign {\smallskip} 
$dF_{occ,NB2090}$                                             &  $1560 \pm 140$                                         \\ \noalign {\smallskip} 
$c1_{I+z}$                                                            & $0.983 \pm 0.050$                                     \\ \noalign {\smallskip}  
$c2_{I+z}$                                                            & $0.065 \pm 0.060$                                     \\ \noalign {\smallskip}                             
$c1_{r'}$                                                               & $1.363 \pm 0.047$                                     \\ \noalign {\smallskip}                            
$c2_{r'}$                                                                & $0.401 \pm 0.051$                                    \\ \noalign {\smallskip}                            
\hline \noalign {\smallskip}
$Deduced$ $stellar$ $parameters$   &    \\ \noalign {\smallskip}
\hline \noalign {\smallskip}
$u1_{I+z}$                                                             & $0.406 \pm 0.026$                 \\ \noalign {\smallskip} 
$u2_{I+z}$                                                             & $0.171 \pm 0.024$                 \\ \noalign {\smallskip} 
$u1_{r'}$                                                                & $0.625 \pm 0.024$                 \\ \noalign {\smallskip} 
$u2_{r'}$                                                                & $0.112 \pm 0.020$                 \\ \noalign {\smallskip} 
$V_\gamma$ [km s$^{-1}$]                                   & $-3.5950 \pm 0.0040$            \\ \noalign {\smallskip} 
Density $\rho_\star$  [$\rho_\odot $]                    & $2.410_{-0.075}^{+0.079}$    \\ \noalign {\smallskip} 
Surface gravity $\log g_\star$ [cgs]                      & $4.645_{-0.010}^{+0.011}$    \\ \noalign {\smallskip} 
Mass $M_\star $    [$M_\odot$]                            & $0.717 \pm 0.025$                 \\ \noalign {\smallskip} 
Radius  $ R_\star $   [$R_\odot$]                         & $0.667_{-0.011}^{+0.010}$     \\ \noalign {\smallskip} 
$T_\mathrm{eff}$ [K]$^a$                                    & $4520 \pm 120$                     \\ \noalign {\smallskip} 
[Fe/H] [dex]$^a$                                                   & $-0.01 \pm 012$                     \\ \noalign {\smallskip}  
 \hline \noalign {\smallskip}
$Deduced$ $planet$ $parameters$   &    \\ \noalign {\smallskip}
\hline \noalign {\smallskip}
RV $K$ [\ms]                                                        & $547.9_{-5.4}^{+5.5}$                                               \\ \noalign {\smallskip} 
$R_p/R_\star$                                                      & $0.15945_{-0.00077}^{+0.00076}$                            \\ \noalign {\smallskip} 
$b_{tr}$ [$R_\star$]                                               & $0.6580_{-0.0095}^{+0.0089}$                                  \\ \noalign {\smallskip} 
$b_{oc}$ [$R_\star$]                                             & $0.655_{-0.013}^{+0.012}$                                         \\ \noalign {\smallskip} 
$T_{oc}-2450000$ [$BJD_{TDB}$]                      &  $5726.95069_{-0.00078}^{+0.00084}$                       \\ \noalign {\smallskip} 
Orbital semi-major axis $a$ [AU]                         &  $0.01526 \pm 0.00018$                                              \\ \noalign {\smallskip} 
Roche limit $a_R$ [AU]                                        &  $0.00768 \pm 0.00016$                                              \\ \noalign {\smallskip} 
$a/a_R$                                                                &  $1.986_{-0.029}^{+0.030}$                                         \\ \noalign {\smallskip} 
$a / R_\star$                                                         & $4.918_{-0.051}^{+0.053}$                                         \\ \noalign {\smallskip} 
Orbital inclination $i_p$ [deg]                               & $82.33 \pm 0.20$                                                        \\ \noalign {\smallskip} 
Orbital eccentricity $ e $                                       & $0.0035_{-0.0025}^{+0.0060}$, $<0.0298$ (3-$\sigma$) \\ \noalign {\smallskip}
Argument of periastron  $ \omega $ [deg]            & $-32_{-34}^{+115}$                                                      \\ \noalign {\smallskip} 
Equilibrium temperature $T_{eq}$ [K]$ $$^b$      & $1440_{-39}^{+40}$                                                    \\ \noalign  {\smallskip} 
Density  $ \rho_p$ [$\rho_{\rm Jup}$]                    &  $1.826_{-0.078}^{+0.084}$                                         \\ \noalign {\smallskip} 
Density  $ \rho_p$ [$g/cm^3$]                              &  $1.377_{-0.059}^{+0.063}$                                         \\ \noalign {\smallskip} 
Surface gravity $\log g_p$ [cgs]                           &  $3.672_{-0.012}^{+0.013}$                                          \\ \noalign  {\smallskip} 
Mass  $ M_p$ [$M_{\rm Jup}$]                             & $2.034_{-0.051}^{+0.052}$                                           \\ \noalign {\smallskip} 
Radius  $ R_p $ [$R_{\rm Jup}$]                          & $1.036 \pm 0.019$                                                       \\ \noalign {\smallskip} 
\hline \noalign {\smallskip}
\end{tabular}
\caption{Median and 1-$\sigma$ limits of the posterior marginalized PDFs obtained for the WASP-43 
system derived from our global MCMC analysis. $^a$From stellar evolution modeling (Sect. 3.3).
 $^b$Assuming $A$=0 and $F$=1. }
\end{center}
\end{table}

\subsubsection{Global analysis of the 23 transits with free timings}

As a complement to our global analysis of the whole data set, we performed a global analysis of the 23 transit light curves alone. 
The goal here was to benefit from the strong constraint brought on the transit shape by the 23 transits  to derive more accurate
transit timings  and to assess the periodicity of the transit. In this analysis, we kept fixed the parameters $T_0$ and $P$ to the 
values shown in Table 5, and we added a timing offset as jump parameter for each transit. The orbit was assumed to be circular. 
The other jump parameters were $dF$, $W$, $b'$, and the limb-darkening coefficients $c_1$ and $c_2$ for both filters. The 
resulting transit timings and their errors are shown in Table 6. Fitting a transit ephemeris by linear regression with these new
derived transit timings, we obtained $T(N) = 2,455,528.868227 (\pm 0.000078) + N \times 0.81347764 (\pm 0.00000065)$ 
BJD$_{TDB}$. Table 6 shows (last column) the resulting transit timing residuals (observed minus computed, O-C). Fig.~5 
(lower panel) shows them as a function of the transit  epochs. The scatter of these timing residuals is now 1.8 times larger than the 
mean error, 19s. Comparing the errors on the timings derived from individual and global analysis of the transit photometry, one 
can notice that the better constraint on the transit shape for the global analysis improves the error of a few timings, but that most
of them have a slightly larger error because of the better separation of the  actual transit signal from the red noise (see above).

\begin{table}
\begin{center}
\begin{tabular}{cccc}
\hline \noalign{\smallskip} 
Epoch & Filter &  $T_{tr}$                              & O-C               \\
           &         &  [BJD$_{TDB}$-2450000]    & [min]               \\     \noalign{\smallskip}  
\hline \noalign{\smallskip} 
11 & $I+z$       & $5537.81673_{-0.00046}^{+0.00047}$  & $0.36 \pm 0.68$ \\ \noalign {\smallskip} 
22 & $I+z$       &  $5546.76494 \pm 0.00022$                  & $0.30 \pm 0.32$  \\ \noalign {\smallskip} 
27 & $I+z$       &  $5550.83219_{-0.00016}^{+0.00015}$ & $0.10 \pm 0.23$  \\ \noalign {\smallskip} 
38 & $I+z$       &  $5559.78033 \pm 0.00021$                  & $-0.07 \pm 0.30$ \\ \noalign {\smallskip} 
38 &Gunn-$r'$  & $5559.78078_{-0.00017}^{+0.00016}$  & $0.58 \pm 0.24$  \\ \noalign {\smallskip} 
43 & $I+z$       & $5563.84771 \pm 0.00022$                   & $-0.08 \pm 0.32$ \\ \noalign {\smallskip} 
49  & $I+z$      & $5568.72836 \pm 0.00015$                   & $-0.39 \pm 0.22$ \\ \noalign {\smallskip}
59 & $I+z$       & $5576.86380_{-0.00015}^{+0.00016}$  & $0.57 \pm 0.23$  \\ \noalign {\smallskip} 
65  & $I+z$      & $5581.74410 \pm 0.00020$                   & $-0.25 \pm 0.29$ \\ \noalign {\smallskip}
70 & $I+z$       &  $5585.81205_{-0.00022}^{+0.00021}$ & $0.56 \pm 0.32$  \\ \noalign {\smallskip} 
76 & Gunn-$r'$ & $5590.69259 \pm 0.00018$                  & $0.09 \pm 0.26$   \\ \noalign {\smallskip} 
87 & Gunn-$r'$ & $5599.64043_{-0.00025}^{+0.00026}$  & $-0.51 \pm 0.37$ \\ \noalign {\smallskip}
97 & $I+z$       & $5607.77517 \pm 0.00014$                  & $-0.56 \pm 0.20$  \\ \noalign {\smallskip}
124 & $I+z$     & $5629.73995_{-0.00018}^{+0.00019}$ &$0.71 \pm 0.27$    \\ \noalign {\smallskip} 
140 & $I+z$     & $5642.75450 \pm 0.00015$                 & $-0.86 \pm 0.22$  \\ \noalign {\smallskip} 
141 & $I+z$     & $5643.56872 \pm 0.00028$                 &  $0.21 \pm 0.40$  \\ \noalign {\smallskip} 
152 & $I+z$     & $5652.51586_{-0.00025}^{+0.00024}$ & $-1.39 \pm 0.36$  \\ \noalign {\smallskip}
168  & $I+z$    & $5665.53229_{-0.00018}^{+0.00019}$ & $-0.26 \pm 0.27$  \\ \noalign {\smallskip}
173 &  $I+z$    & $5669.59976 \pm 0.00023$                  & $-0.14 \pm 0.33$  \\ \noalign {\smallskip}
189 &  $I+z$    & $5682.61584_{-0.00023}^{+0.00022}$ &$0.49 \pm 0.33$    \\ \noalign {\smallskip}
200 &  $I+z$    & $5691.56383_{-0.00023}^{+0.00022}$ & $0.11 \pm 0.33$  \\ \noalign {\smallskip}
211  & $I+z$    & $5700.51237_{-0.00015}^{+0.00014}$ & $0.52 \pm 0.22$  \\ \noalign {\smallskip}
243  & $I+z$    & $5726.54399 \pm 0.00035$                  & $1.00 \pm 0.50$  \\ \noalign {\smallskip}
\hline
\end{tabular}
\caption{Median and 1-$\sigma$ errors of the posterior PDFs  deduced for the timings of the transits
from their global analysis. The last column shows the difference (and its error) between the measured timing 
and the one deduced from the best-fitting transit ephemeris computed by linear regression, 
$T(N) = 2455528.868227 (0.000078) + N\times 0.81347764 (0.00000065)$ BJD$_{TDB}$.  }
\end{center}
\label{wasp43-TTV}
\end{table}

\section{Discussion}

\subsection{The physical parameters of WASP-43\,b}

Comparing our results for the WASP-43 system with the ones presented in H11, we notice that our derived 
parameters agree well with the second solution mentioned in H11, while being significantly more precise. 
 WASP-43\,b is thus a Jupiter-size planet, twice more massive than our Jupiter, orbiting at 
only $\sim$0.015 AU from a 0.72 $M_\odot$  main-sequence K-dwarf. Despite having the smallest
orbital distance among the hot Jupiters, WASP-43\,b is far from being the most irradiated known 
exoplanet, because of the small size and temperature of its host star. Assuming a solar-twin host star, 
its incident flux $\sim$9.6 10$^8$ erg~s$^{-1}$~cm$^{-2}$ would correspond to a $P=2.63$ d orbit 
($a = 0.0374$ AU), i.e. to a rather typical hot Jupiter. With a radius of 1.04 $R_{Jup}$, WASP-43\,b lies
 toward the bottom of the envelope described by the published planets in the $R_{p}$ vs. incident flux 
 plane (Fig.~9). Taking into account its level of irradiation, the high density of the planet favors an old age 
 and a massive core under the planetary structure models of Fortney et al. (2010). Our results are 
 consistent with a circular orbit, and we can put a 3-$\sigma$ upper limit $<$0.03 to the orbital eccentricity.
 Despite its very short period, WASP-43\,b has a semi-major axis approximatively twice larger than its Roche 
 limit $a_R$, which is typical for hot Jupiters (Ford \& Rasio 2006, Mastumura et al. 2010). As noted by these authors, 
 the inner edge of the distribution of most hot Jupiters at $\sim$2 $a_R$ favors migrational mechanisms based on the 
 scattering of planets on much wider orbit and the subsequent tidal shortening and circularization of their orbits. 
 Thus, WASP-43\,b does not appear to be a planet exceptionally close to its final tidal disruption, unlike WASP-19\,b 
 that orbits at only 1.2 $a_R$ (Hellier et al. 2011a). 
 
\begin{figure}
\label{fig:9}
\centering                     
\includegraphics[width=9cm,angle=0]{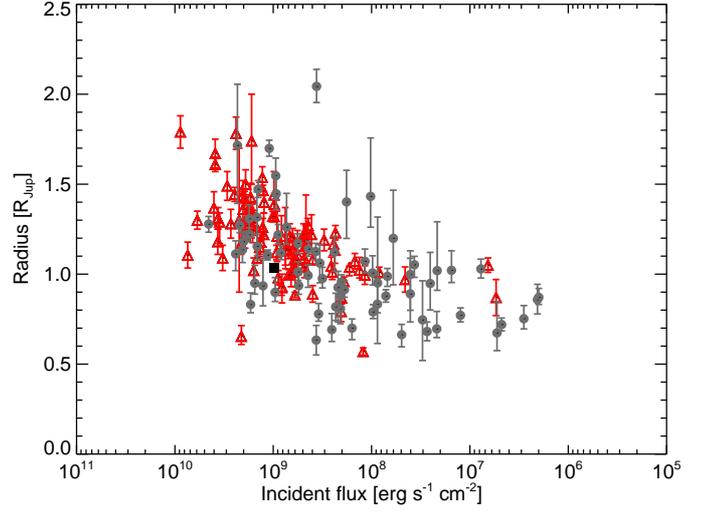}
\caption{Planetary radii as a function of incident flux. WASP-43\,b is shown as a black square. Grey filled 
circles are \textit{Kepler} \textit{planetary} candidates (see Demory \& Seager 2011). Transiting giant planets 
previously published, and mostly from ground-based surveys, are shown as grey triangles. The relevant 
parameters $R_p$, $R_s$, $T_{eff}$ and $a$ have been drawn from http://www.inscience.ch/transits on 
August 29, 2011.}
\end{figure}

\subsection{The atmospheric properties of WASP-43\,b}

We have modeled the atmosphere of WASP-43\,b using the methods described in Fortney et al. (2005, 2008) 
and Fortney \& Marley (2007).  In Fig.~10, we compare the planet-to-star flux ratio data to three atmosphere models.  
The coldest model (orange) uses a dayside incident flux decreased by 1/2 to simulate the loss of half of the absorbed 
flux to the night side of the planet.  Clearly the planet is much warmer than this model.  In blue and red are two models 
where the dayside incident flux is increased by a factor of 4/3 to simulate zero redistribution of absorbed flux (see, e.g., 
Hansen 2008).  The red model features a dayside temperature inversion due to the strong optical opacity of TiO and VO 
gases (Hubeny et al. 2003, Fortney et al. 2006).  The blue model is run in the same manner as the red model, but 
TiO and VO opacity are removed.

The blue and red models are constructed to maximize the emission from the dayside of the planet.  Figure 10 shows that 
only such bright models could credibly match the data points.  While a dayside with no temperature inversion is slightly 
favored by the 1.2 $\mu$m data point, it is difficult to come to a firm conclusion.  Day-side emission measurements from 
{\it Warm Spitzer} will help to better constrain the atmosphere as well. They will also allow getting an even smaller
upper-limit on the orbital eccentricity.

Fortney et al. model fits to near infrared photometry of other transiting planets (e.g., Croll et al. 2010) have 
generally favored inefficient temperature homogenization between the day and night hemispheres, although the warm 
dayside of WASP-43\,b appears to be at the most inefficient end of this continuum.  The apparent effeciency of temperature 
homogenization is expected to vary with wavelength.  In particular, near infrared bands, at minima 
in water vapor opacity, are generally expected to probe deeper into atmospheres than the {\it Spitzer} bandpasses.

\begin{figure}
\label{fig:10}
\centering                     
\includegraphics[width=9cm,angle=0]{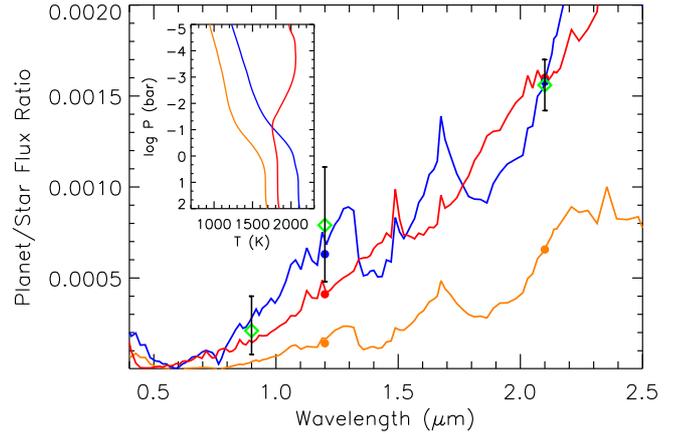}
\caption{Model planet-to-star flux ratios compare to the three data points.  The data are green diamonds with 
1$-\sigma$ error bars shown.  The orange model assumes planet-wide redistribution of absorbed flux.  The 
red and blue models assume no redistribution of absorbed flux, to maximize the day-side temperature.  The 
red model includes gaseous TiO and VO and has a temperature inversion.  [See the figure inset.]  The blue and 
orange models have TiO and VO opacity removed, and do not have a temperature inversion. For each model, filled
circles are model fluxes averaged over each bandpass.  The data slightly favor a model with no day-side temperature 
inversion.}
\end{figure}

\subsection{Transit timings}

Dynamical constraints can be placed on short orbits companion planets from the 23 transits obtained in this study.
The linear fit to the transit timings described above yields a reduced $\chi^2$ of 4.6 and the $rms$ of its residuals
 is 38s. Two transits (epochs 140 and 152) have a O-C different from zero at the $\sim$4-$\sigma$ level. The most 
 plausible explanation for the significant scatter observed in the transit timings is systematic errors on the derived
 timings due to the influence of correlated noise. Another potential explanation is  asymmetries in the transit
  light curves caused by the crossing of one or several star spots by the planet. In such cases, the fitted transit profile is 
  shifted with time, producing timing variations. 
  
We also explored the detectability domain of a second planet in the WASP-43 system. To this end, we followed
 Agol et al. (2005) and used the \textsc{Mercury} n-body integrator package (Chambers 1999). We simulated 3-body
  systems including a second companion with orbital periods ranging between 1.3 and 50 days, masses from 0.1 
  $M_{\oplus}$ to 2.0 $M_{Jup}$ and an orbital eccentricities of $e_c=0$ and $e_c=0.05$. For each simulation, we computed the 
 $rms$ of the computed transits timings variations. The results are shown on Fig.11 where the computed 1-$\sigma$ (red)
 and 3-$\sigma$ (black) detection thresholds are plotted for each point in the mass-period plane. We also added RV detectability
threshold curves based on 5 (solid), 10 (dash) and 15 (dot) \ms semi-amplitudes ($K$).

Based on the present set of timings, the timings variations caused by a 5 Earth mass companion in 2:1 resonance 
would have been detected with 3-$\sigma$ confidence, while unseen with radial velocities alone.

\begin{figure}
\label{fig:11}
\centering                     
\includegraphics[width=8cm,angle=0]{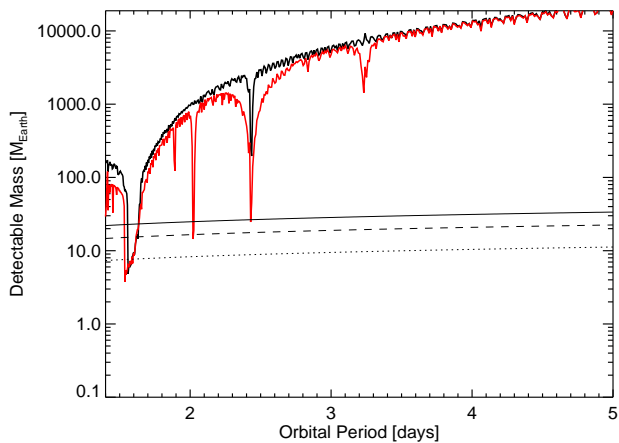}
\caption{Detectivity domain for a putative WASP-43\,c planet, assuming $e_c=0$ (black) and $e_c=0.05$ (red). 
The solid curves delimit the mass-period region where planets yield maximum TTV on WASP-43\,b above 114 s 
(3-$\sigma$ detection based on the present data). The dotted curves show the 1-$\sigma$ threshold. Nearly 
horizontal solid, dashed and dotted lines shows RV detection limits for RV semi-amplitude $K$=5, 10 and 15 \ms  
respectively. }
\end{figure}

\section{Conclusions}

In this work we have presented 23 transit light curves and 7 occultation light curves for the ultra-short 
period planet WASP-43\,b. We have also presented 8 new measurements of the radial velocity of the star.
Thanks to this extensive data set, we have significantly improved the parameters of the system. Notably, 
our strongly improved precision on the stellar density ($2.41 \pm 0.08 \rho_\odot$) combined with a very 
reasonable constraint on its age (to be younger than a Hubble time) allowed us to break the degeneracy of
 the stellar solution mentioned by H11. The resulting stellar mass and size are $0.717 \pm 0.025 M_\odot$ 
 and $0.667 \pm 0.011 R_\odot$.  Our deduced physical parameters for the planet are $2.034 \pm 0.052 
 M_{Jup}$ and $1.036 \pm 0.019 R_{Jup}$. Taking into account its level of irradiation, the high density of the
  planet favors an old age and a massive core. Our deduced orbital eccentricity, $0.0035_{-0.0025}^{+0.0060}$, 
is consistent with a fully circularized orbit. 
  
The parameters deduced from the individual analysis of the 23 transit light curves show some extra scatter that
we attribute to the correlated noise of our data and, possibly, to the crossing of spots during some transits. 
This conclusion is based on the correlation observed among the transit parameters. These results reinforce 
the interest of performing global analysis of   extensive data sets in order to minimize systematic errors and 
to reach high accuracies on the derived parameters. 

We detected the emission of the planet at 2.09 $\mu$m at better than 11-$\sigma$, the deduced occultation
depth being $1560 \pm 140$ ppm. Our detection of the occultation at 1.19 $\mu$m is marginal ($790 \pm 320$ ppm)
and more observations would be needed to confirm it. We place a 3-$\sigma$ upper limit of 850 ppm on the depth
of the occultation at $\sim$0.9 $\mu$m. Together, these results strongly favor a poor  redistribution of the heat
from the dayside to the nightside of the planet, and marginally favor a model with no day-side temperature
inversion. 

\begin{acknowledgements}
TRAPPIST is a project funded by the Belgian Fund for Scientific Research (Fond National de la Recherche Scientifique, 
F.R.S-FNRS) under grant FRFC 2.5.594.09.F, with the participation of the Swiss National Science Fundation (SNF).  M. 
Gillon and E. Jehin are FNRS Research Associates. We are grateful to ESO La Silla and Paranal staffs for their 
continuous support. We thank the anonymous referee for his valuable suggestions. 
\end{acknowledgements} 

\bibliographystyle{aa}

\end{document}